%% file: ms.tex
\documentclass[]{article}
\usepackage{url}
\usepackage{eqparbox}
\usepackage{multirow}
\usepackage{subcaption}
\usepackage{amsmath}
\usepackage{siunitx}
\usepackage[colorinlistoftodos]{todonotes}
\usepackage{algorithm}
\usepackage[noend]{algpseudocode}
\usepackage{array}
\usepackage{float}
\usepackage{verbatim}
\usepackage{hyperref}
\usepackage{xspace}
\usepackage{enumitem}
\usepackage{hyperref}
\usepackage[capitalise]{cleveref}
\usepackage[a]{esvect}       
\usepackage{footmisc}
\usepackage[export]{adjustbox}
\makeatletter
\let\ftype@table\ftype@figure
\makeatother

\makeatletter
\algnewcommand{\LineComment}[1]{\Statex \hskip\ALG@thistlm \(\triangleright\) #1}
\makeatother

\newcommand{\cac}{\textsc{CAC\xspace}}
\newcommand{\hypa}{HYPA\xspace}

\title{Sequential Motifs in Observed Walks}
\author{Timothy LaRock\thanks{Network Science Institute, Northeastern University, Boston, MA, USA}~~~~ Ingo Scholtes\thanks{Data Analytics Group, University of Z\"urich. Z\"urich, Switzerland}~\thanks{Chair of Machine Learning for Complex Networks, Center for Artificial Intelligence and Data Science (CAIDAS), Julius-Maximilians-Universität Würzburg, Germany}~~~~ Tina Eliassi-Rad$^*$\thanks{Khoury College of Computer Sciences, Northeastern University, Boston, MA, USA}}

\begin{document}

\maketitle

\begin{abstract}
The structure of complex networks can be characterized by counting and analyzing network \emph{motifs}.
Motifs are small subgraphs that occur repeatedly in a network, such as triangles or chains.
Recent work has generalized motifs to temporal and dynamic network data.
However, existing techniques do not generalize to \emph{sequential} or \emph{trajectory} data, which represents entities moving through the nodes of a network, such as passengers moving through transportation networks. 
The unit of observation in these data is fundamentally different, since we analyze full observations of trajectories (e.g., a trip from airport A to airport C through airport B), rather than independent observations of edges or snapshots of graphs over time.
In this work, we define \emph{sequential motifs} in trajectory data, which are small, directed, and edge-weighted subgraphs corresponding to patterns in observed sequences.
We draw a connection between counting and analysis of sequential motifs and Higher-Order Network (HON) models. We show that by mapping edges of a HON, specifically a  $k$th-order DeBruijn graph, to sequential motifs, we can count and evaluate their importance in observed data.
We test our methodology with two datasets: (1) passengers navigating an airport network and (2) people navigating the Wikipedia article network. We find that the most prevalent and important sequential motifs correspond to intuitive patterns of traversal in the real systems, and show empirically that the heterogeneity of edge weights in an observed higher-order DeBruijn graph has implications for the distributions of sequential motifs we expect to see across our null models.
\end{abstract}

\maketitle
\setcounter{footnote}{0}
\section{Introduction}
\label{sec:intro}
Motifs in complex networks are small subgraphs that can be counted and analyzed from observed data. Methods for using motifs to characterize static complex networks have been developed over the last 20 years \cite{milo2002network, artzy-randrup2004comment, saramaki2005characterizing, underwood2020motifbased}, and equivalent or similar concepts were investigated even earlier in the social sciences (see e.g., triadic analysis as reviewed in \cite{wasserman1994social}). More recent research has focused on identifying motif structures in \emph{temporal} or \emph{dynamic} network data, which comes in the form of either timestamped independent edge observations or snapshots of a process over time \cite{kovanen2011temporal, jurgens2012temporal, kovanen2013temporal, xuan2015temporal, paranjape2017motifs, tu2018network, liu2020temporal}. In this work, we focus our attention on data that represents \emph{sequences}, \emph{trajectories}, or \emph{walks} through networks.\footnote{We will consider these terms interchangeable.} Observations in this data are walks of varying lengths through a network. Each walk is observed independently and generally without temporal information about the appearance of individual edges within the walk. These walks may represent trajectories through physical networks (for example passengers or freight moving through a railway network) or through non-physical networks (for example sequences of proteins that make up the proteome of an organism, or sequences of words in a language such as n-grams \cite{pibiri2019handling}).

The goal of this work is to count and analyze \emph{sequential motifs} in observed walks. Intuitively, a sequential motif is a small directed subgraph that is an abstract representation of a walk through a network. In the simplest case, we can think of an edge from node A to node B, or the self-loop from node A to itself, as sequential motifs of length 1 edge. These structures are the objects of study in traditional network analysis, thus in practice we exclude them from our analyses; and in general, we will exclude self-loops. At length 2 edges, there are two motifs: the backtracking or bidirectional motif A-B-A, and the chain motif A-B-C. These motifs are both relevant for understanding the interaction between a network structure and the processes that unfold on top of it. For example, in an airport network representing passenger movements, the sequential motif A-B-A represents a round-trip flight originating and terminating at the same airport. 

\begin{figure}[!ht]
		\hspace{-1cm}\includegraphics[scale=0.3]{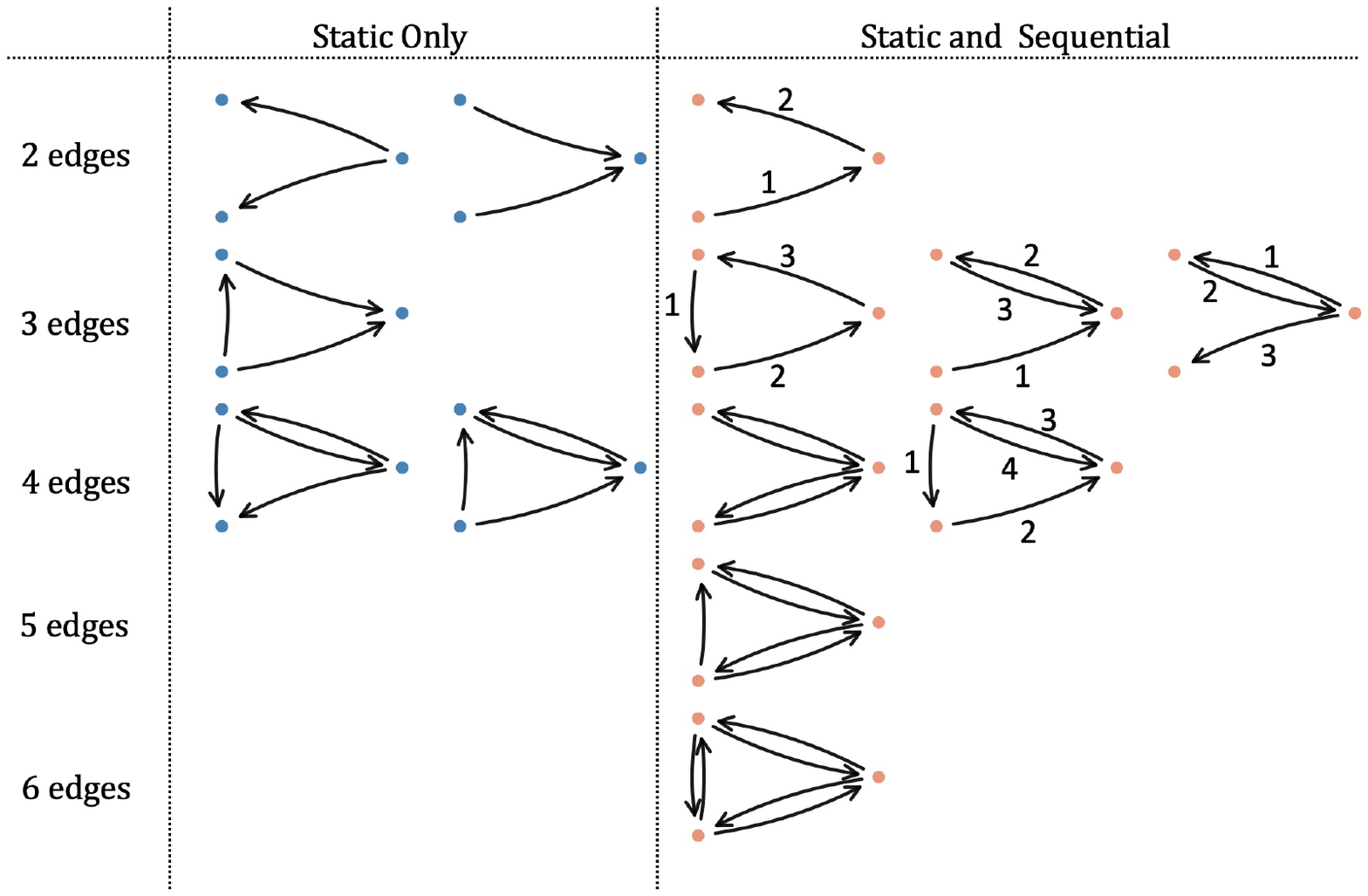}%
		\hspace{0.2cm}\includegraphics[scale=0.45]{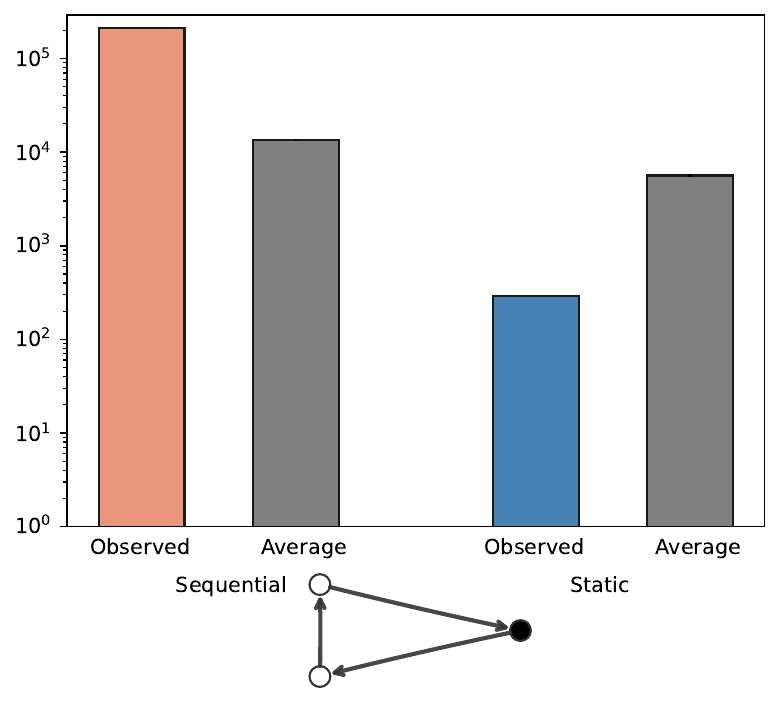}
	\caption{
		Left: Enumeration of 3-node simple directed motifs. Motifs in the right column have interpretations in both sequential data and static graphs, while those in the left column have no sequential interpretation because the edges involved cannot appear in sequence (equivalently there is no \emph{Eulerian path} through the motif starting from any node).
		Right: Comparison of the frequency of the directed cycle motif on 3-nodes (i.e., a directed triangle) in data representing passenger trajectories through the airport-to-airport network. The first (orange) bar shows the  observed frequency using the proposed sequential motif methodology. The second bar shows the same  count but averaged over many randomizations of the data. The third (blue) bar shows the observed count in a static, directed, and unweighted graph, computed using the G-tries algorithm \cite{ribeiro2014gtries}. The last bar again shows the average frequency of the motif in randomized networks. Static motif analysis suggests cycles are under-represented, since the motif is more prevalent in randomized networks. In contrast, sequential motif analysis shows that the directed triangle is over-represented in the motifs compared with random realizations of the DeBruijn graph ensemble.
		}
	\label{fig:static_vs_seq}
\end{figure}

In the left panel of \cref{fig:static_vs_seq}, we show all simple (e.g., no multi-edges) directed motifs on 3-nodes. The left column shows motifs that are only interpretable in static network data, while the right column shows motifs that can be interpreted in either static or sequential contexts. A motif has a sequential interpretation if, starting at any node, there is a \emph{directed Eulerian path}--a path that visits each edge exactly once--through the simple (i.e., without parallel edges) version of the motif. Motifs without Eulerian paths cannot possibly correspond to observed sequences of edges, since the lack of such a path indicates that there is at least 1 edge that cannot be traversed in sequence.\footnote{Note that the motif itself does not have to be a Eulerian path, since motifs may contain parallel edges. For example, A-B-A-B is a valid sequential motif, but is not itself a Eulerian path, since the edge A-B is repeated.} In the right panel of \Cref{fig:static_vs_seq}, we show the observed frequency of the directed triangle sequential motif in data representing passenger flight itineraries (\cite{transstat2018origin}, see \cref{sec:results} for details). For air travel, the cycle motif corresponds to a trip with a direct flight in one direction and a layover in the other, or a trip that visits two different locations before returning to the origin, such as a multi-city business trip. The key takeaway from this comparison is that the directed triangle appears to be \emph{overrepresented} compared to random expectation of sequential motifs, since it is observed considerably more often in the data than in the randomized DeBruijn graphs. However, based on static motifs computed from a directed and unweighted graph constructed from the flights (using the G-tries algorithm \cite{ribeiro2014gtries}), the motif appears to be underrepresented, as it occurs slightly less frequently than expected at random. This sort of discrepancy is the motivation for our work.

Our methodology is strongly connected to recent research on Higher-Order Network Models (HONs). Here, higher-order refers to sequential and temporal correlations in how a network is traversed that violate the typical Markovian assumption implicit in the study of traditional, first-order networks \cite{scholtes2014causalitydriven, scholtes2016higherorder, xu2016representing, scholtes2017when, lambiotte2019networks, larock2020hypa}.\footnote{This is in contrast to research on \emph{higher-order structures} or \emph{polyadic interactions}, such as hypergraphs or simplicial complexes \cite{chodrow2020configuration, battiston2020networks, torres2021why}.} The violated assumption is that the $t+1$st step taken by a walker moving randomly on the edges of a network is dependent only on the position of the walker at time $t$, e.g., that the walker has no \emph{memory} of where it was at times $i<t$. HONs for sequential data directly incorporate memory into a graphical representation, moving the Markovian assumption from edges between two nodes to paths through $k$ nodes. These graphical representations can be then interpreted and analyzed using modified techniques from graph theory and network science. 

In this paper, we count sequential motifs in observed data and evaluate their importance by mapping the edges of a particular HON, known as the \emph{DeBruijn graph}, to their corresponding motif structures. In a DeBruijn graph of order $k$, each node represents a walk of length $k-1$ (e.g., $A\rightarrow B$ for $k=2$); and each directed, weighted edge represents the frequency of a length $k$ walk through a traditional (or first-order) graph (e.g., $A\rightarrow B, B\rightarrow C$).  Our work extends recent research showing the utility of DeBruijn graphs as representations for modeling correlations in pathway data through networks to study motif structures \cite{scholtes2017when, larock2020hypa, gote2020predicting}.\footnote{We note that our methodology can be applied to other HONs, for example the variable-order HON developed in \cite{xu2016representing}. We leave the investigation of motifs using these models for future work.}

\begin{figure}[ht]
	\centering\includegraphics[scale=0.5]{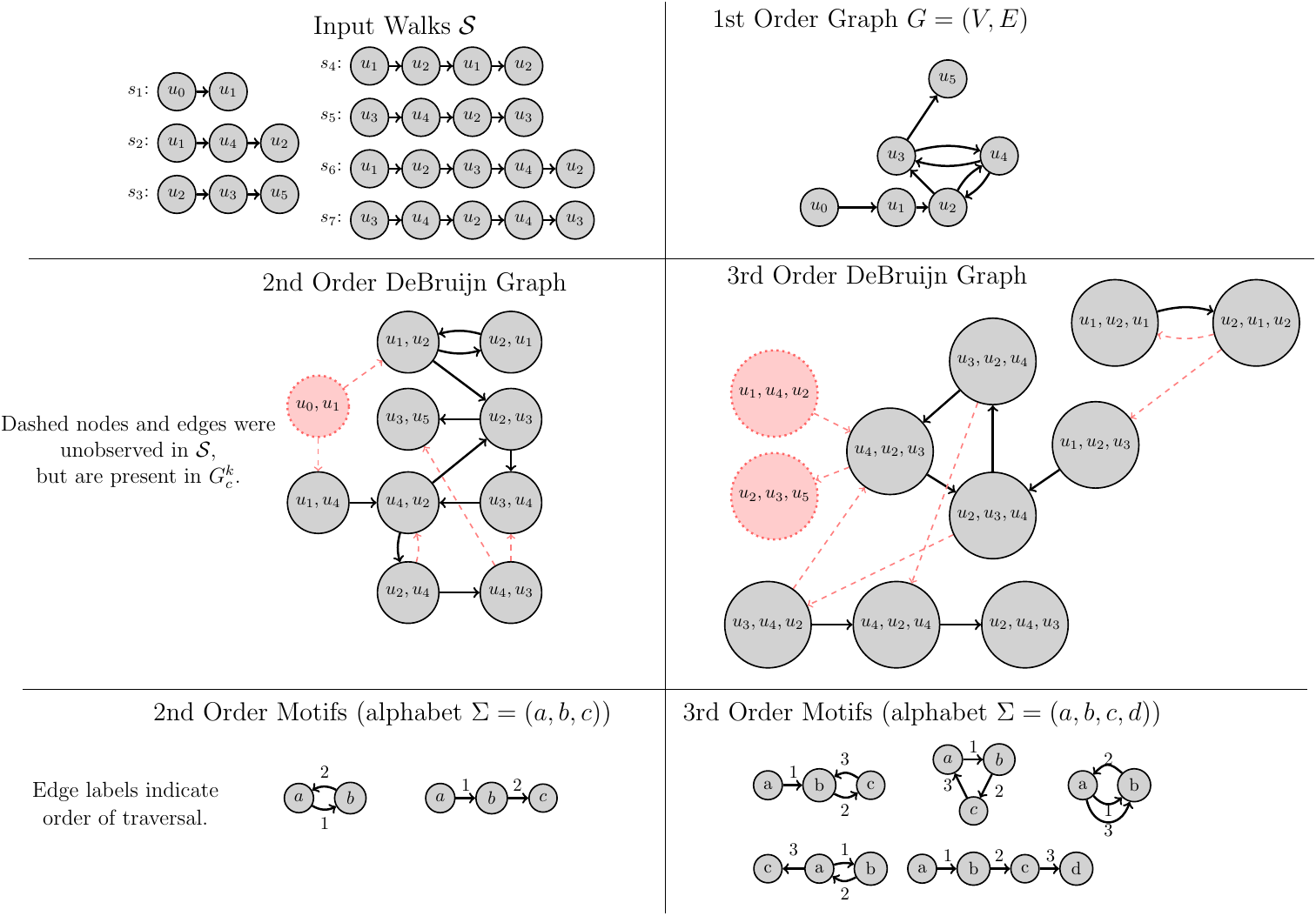}
	\caption{Example illustrating the relationship between walks, DeBruijn graphs, and sequential motifs. The top left panel shows the input dataset $\mathcal{S}$ consisting of 7 walks on nodes $\{u_1, u_2, u_3, u_4, u_5\}$. The top right panel shows the first-order directed graph $G$ constructed from the input data. The middle panels show DeBruijn graphs of order 2 and 3. Solid lines indicate the observed DeBruijn graph, while dashed lines indicate unobserved nodes and edges that exist in the complete DeBruijn graph $G^k_c$. The bottom panels show the sequential motifs counted from $\mathcal{S}$ for orders 1 and 2. Labels on the edges of the motif indicate the order in which the edges are traversed.}
	\label{fig:example}
\end{figure}

\paragraph{Definitions} 
We compute sequential motifs from a dataset of $n$ walks $\mathcal{S} = \{(s_1, w_1), (s_2, w_2), \cdots (s_n, w_n)\}$. Each $s_i=\langle u_1, u_2, \cdots, u_{|s_i|}\rangle$ represents an $|s_i|$-length walk through a directed graph $G=(V,E)$ with $N=|V|$ nodes and $M=|E|$ edges. For each walk $s_i$, the optional value $w_i$ represents the \emph{frequency} of the observed walk in the data.\footnote{For example, we may observe that 100 passengers bought the same round trip ticket $s_j=\langle u_1, u_2, u_1\rangle$, and thus the observation $s_j$ will have frequency $w_j=100$.}  If frequencies are not provided, $\mathcal{S}$ is considered a multiset where duplicate appearances of the same walk are aggregated and assigned a frequency equal to their multiplicity in $\mathcal{S}$. We denote as $\mathcal{M}^k$ the set of $k$-edge sequential motifs, where each motif $m=\langle \sigma_1, \sigma_2, \cdots, \sigma_{k+1} \rangle \in\mathcal{M}^k$ represents a walk of length $k$ edges through an arbitrary alphabet $\Sigma=\{\sigma_1, \sigma_2, \dots \sigma_{\ell}\}$, where $\ell \leq k+1$ is the maximum number of unique nodes in a $k$-edge walk. The alphabet may consist of any arbitrary set of symbols. The only requirement is that the number of unique symbols is at least as great as the number of unique nodes in any observed walk. Therefore, the worst-case alphabet size is the number of nodes $N$.\footnote{In principle we can compute sequential motifs of any length $k$, provided that sequences of length at least $k$ edges exist in the input data. However, larger values of $k$ quickly become impractical to analyze. Therefore, we we will limit our results to motifs up to 3 edges, with the exception of \Cref{fig:static_vs_seq}, where we show motifs up to 6-edges.} Since we assume our input data are walks through a network, sequential motifs $m\in\mathcal{M}^k$ may have parallel edges that represent multiple traversals of the same edge in $m$.\footnote{We could equivalently say the edges of sequential motifs are weighted by the number of times the edge is used. However, to avoid confusion with other notions of ``weight" used in later analyses, we describe motifs as multigraphs for clarity.} Finally, we define a $k$th-order DeBruijn graph as $G^k=(V^k,E^k)$ where $V^k$ is the set of $N_k$ $k$th-order nodes $\vv{v}$, each representing a path of length $k-1$ through $G$, and $E^k$ is the set of $M_k$ $k$th-order edges $(\vv{v},\vv{w})$, each representing a path of length $k$. The higher-order nodes in an edge $(\vv{v},\vv{w})\in E^k$ must overlap in $k-1$ first-order nodes, i.e., $\vv{v}[1,\cdots,k] = \vv{w}[0,\cdots,k-1]$ for all ${(\vv{v},\vv{w})\in E^k}$. 

\paragraph{Contributions}
Our main contribution is to make explicit the connection between sequential motifs and DeBruijn graph models of observed walks. We use this connection to develop methods for evaluating the \emph{importance} of a given sequential motif to a dataset or process (see \cref{subsec:importance}). Since the edges of DeBruijn graphs correspond to walks of length $k$, we can interpret randomization of the DeBruijn graph structure as randomization of walks of length $k$ (i.e., motif realizations, and vice versa). Thus our methods attempt to answer the question: \emph{is this motif realized substantially more or less often in the observed data than expected at random?}

The key insight driving our approach is that we can count sequential motifs involving $k$ edges by constructing a $k$th-order DeBruijn graph $G^k$ and for every edge $e=(\vv{v},\vv{w})\in E^k$ that defines a $k$-edge walk $p_e = \vv{v} + \vv{w}[k-1]$ mapping the first-order nodes $u_i \in p_e, i\in 1\dots k+1$ one-to-one with the alphabet $\Sigma$. Projecting every edge in this way and counting the frequency of each projected motif pattern allows us to uncover patterns of traversal in observed walks. \cref{fig:example} shows a toy example illustrating this process. We then adopt and extend appropriate null models for DeBruijn graphs, which we use to evaluate the statistical importance of the motifs based on different randomizations.

Our work makes the following contributions to the literature on understanding patterns in trajectories through complex networks:
\begin{itemize}
	\item We define \emph{sequential motifs}, which are small and directed subgraphs, sometimes with parallel edges, representing abstractions of walks through a directed graph.
	\item We show how counting sequential motifs can be achieved by mapping the edges of a DeBruijn graph to motifs; and provide an algorithm called \cac\ for simultaneously constructing a DeBruijn graph and counting sequential motifs in time that scales linearly with the number of observed walks.
	\item We provide methods for evaluating the importance of motifs based on sampling edges uniformly at random from complete DeBruijn graphs, which represent all possible walks of length $k$ edges.
	\item We study sequential motifs in datasets of passenger trips through the domestic airport network in the United States, as well as clickstreams representing players navigating from random source pages to random targets in Wikipedia. We find correspondence between how we expect people to navigate the networks and the motifs that appear more or less often than expected at random based on our null models.
\end{itemize}

The remainder of the paper is organized as follows. In the next section, we review some work related to our research. In \cref{sec:method}, we describe our methodology for computing sequential motifs and evaluating their importance. Then, in \cref{sec:results}, we present empirical experiments and discuss the circumstances in which sequential motifs are the right tool for analyzing a dataset. Finally, we conclude the paper in \cref{sec:conclusion} with a discussion of future directions for this research.

\section{Related Work}
\label{sec:relatedwork}

Motifs in network data have been studied in various forms across disciplines for decades \cite{stone2019network, ribeiro2019survey, jazayeri2020motif}. The formulation presented in this work was inspired by work published mostly in the last twenty years, starting with the identification of network motifs as building blocks of complex networks \cite{milo2002network}. This work measured and compared motifs across contexts, including gene regulatory networks (building on \cite{shen-orr2002network}), neuronal networks, food webs, electronic circuits, and the World Wide Web, using network randomization to define a notion of significance for each motif. Much of the early work on  motifs in biological systems has been reviewed in \cite{alon2007network} and more recently in \cite{patra2020review}. The null models that underly motif comparisons are typically samples from \emph{ensembles} of random graphs that preserve properties relevant to the particular networks under study \cite{hartle2020network}. Some work has also been done investigating \emph{heterogeneous} network motifs, where heterogeneous refers to networks where nodes do not all have the same type \cite{rossi2019heterogeneous}.

The closely related concept of \emph{closed frequent subgraph mining} was developed in parallel in the data mining community \cite{yan2003closegraph, yan2005mining}. In this formulation, the goal is to quickly identify interesting and maximal subgraph patterns in very large graph datasets, including over sets of graphs. These patterns are typically larger and more complex than what is studied in the literature on network motifs.

Another related literature addresses higher-order \emph{structure} in graphs \cite{battiston2020networks}, for example by studying simplicial complices \cite{young2017construction, petri2018simplicial, benson2018simplicial, iacopini2019simplicial} and hypergraphs \cite{benson2019three, chodrow2020configuration}, including work on hypergraph motifs \cite{lee2020hypergraph}. Both of these areas are focused on simultaneous interaction among more than two nodes, sometimes referred to as polyadic interactions \cite{chodrow2020configuration}. In contrast, our focus is on data that is characterized by dyadic interactions that happen in sequence.

Also related is the study of \emph{subgraph frequencies}, most notably in \cite{ugander2013subgraph}. This work defines a coordinate system for collections of disconnected graphs based on the frequency of small connected subgraphs, defined equivalently to motifs.

The work discussed so far largely addresses networks viewed as \emph{static}, meaning that the nodes and edges remain the same over time. Yet another field of study has emerged to understand motifs in \emph{temporal} network data \cite{kovanen2011temporal, jurgens2012temporal, kovanen2013temporal, xuan2015temporal, paranjape2017motifs, tu2018network, liu2020temporal}. Temporal network data comes in the form of timestamped edge observations $e=(u,v,t)$, where $u$ and $v$ are nodes and $t$ is a timestamp. Temporal, dynamic, or streaming network data is common, especially in (online) communication \cite{abello2010detecting, kovanen2011temporal, jurgens2012temporal, kovanen2013temporal, xuan2015temporal, tu2018network} and also in biological systems \cite{sarkar2019using}. The difference between temporal and sequential data is the unit of observation: temporal network data usually consists of independent and timestamped edge observations, while sequential data consists of complete observations of sequences, usually without timestamps. In this data we begin from whole observations of sequences, but we do not necessarily know the order of the observations of the sequences themselves or the subobservations (e.g., individual edges) across different sequences, since we do not have timestamps. In contrast, in temporal network data we only have partial observations (individual edges) and need to infer or define a time-scale to find ``whole'' observations of paths, but we know for certain the order of the individual events given their timestamps. Although we do not address it in the current paper, this distinction does not necessarily impede us from analyzing temporal network data using sequential motifs. We can apply existing techniques to transform edge data into pathway datasets \cite{petrovic2019counting}, then apply our sequential motif analysis to the resulting pathway dataset. This moves the problem of determining the appropriate timescale at which to study the data to a pre-processing step \cite{soundarajan2016generating}.

Our work is also related to research on community detection using the line graph transformation, which is closely related to the DeBruijn graph \cite{evans2009line}. An unweighted 2nd-order DeBruijn graph is the same as the line graph transformation of the first-order representation. In \cite{evans2009line}, the concept of \emph{modularity} that is the basis of many community detection algorithms is generalized to line graphs to discover link partitions or communities in the network.

Our work is also related to the concept of $k$-motifs introduced by Sinatra et al.  \cite{sinatra2010networks}. They identified subsequences of strings of symbols that they described as ``fundamental units" of the process generating the strings, analogous to identifying words in English sentences where the punctuation has been removed. Although our methods seem similar at first glance, there is a fundamental difference in our respective deployments of the word ``motif." In our case, a motif is defined arbitrarily, and trajectories through real networks can realize the abstract motif. Or, put another way, we are interested in motifs as \emph{types} and walks as \emph{tokens}, each of which is a realization of a type. In Sinatra et al \cite{sinatra2010networks}, the most significant realizations--strings representing observed sequences, i.e., tokens--are taken to be motifs, and there is no study of motifs as we define them in this work. In their study, a $k$-motif is defined as a significantly re-occuring sub-sequence in a sequential dataset. They defined and analyzed networks of $k$-motifs by first determining which subsequences were observed at higher than expected frequencies based on statistical significance of the observed frequencies in $k$th order Markov models. They then defined significant $k$-motifs as nodes in a motif graph, and connected pairs of motifs if they co-occurred in at least one sequence of the input data, where two motifs $(i,j)$ co-occur if $i$ appears immediately before $j$. Finally, they filtered out edges between $k$-motifs whose co-occurrences were not significant with respect to random expectation. They go on to show that community detection algorithms on $k$-motif networks can uncover patterns in diverse data, from clusters of proteins in the human proteome, to information cascades on online social media platforms.

Recent work has attempted to understand the role of motifs in \emph{dynamics} on networks \cite{sekara2016fundamental, sarkar2019using, schwarze2020motifs}.  Our work is related to that of Schwarze and Porter \cite{schwarze2020motifs}, who recently defined \emph{process motifs} on graphs. A process motif is defined by the \emph{walks} that are possible on a specific substructure of a network. Schwarze and Porter \cite{schwarze2020motifs} show how the connection between structural and process motifs can be used to determine the role of different substructures in shaping a dynamical process on a network. Our work focuses on counting structural motifs from sequential data, but we note that because sequential data is the result of a discrete dynamical process (e.g., a random walk), our work blends structure and process, suggesting a close connection with this line of research.

We also build on the literature on DeBruijn graphs, which have been studied across fields for decades. In discrete mathematics, for example, the appearance and disappearance of cycles in DeBruijn graphs was a topic of interest in the 1970s \cite{mykkeltveit1972proof, lempel1971extremal}. 
In biology and bioinformatics, variations of DeBruijn graphs are widely used to assemble DNA sequencing data into full genomes \cite{pevzner2001eulerian, zerbino2008velvet, iqbal2012novo, garimella2020detection}. In computer networks, optimal or near-optimal routing schemes have been found for DeBruijn graph representations \cite{bermond1986strategies, loguinov2005graphtheoretic,  chikhi2015representation, faizian2018random}, and DeBruijn graphs have been used to design and analyze feedback shift registers in memory systems \cite{lempel1970homomorphism, chee2020constrained}. Most recently, DeBruijn graphs have been introduced as an appropriate representation for sequential and pathway data on networks \cite{scholtes2017when, larock2020hypa, gote2020predicting}, which is the line of research we are contributing to most directly.

Finally, our research is related to the literature on network sampling. In general, network sampling is the process of gathering random samples of the network, for example random nodes or random edges (for a review, see \cite{ahmed2014network}). These samples can then be used to approximate important network properties like the degree distribution \cite{gkantsidis2006random, yoon2007statistical, costa2007exploring, ribeiro2010estimating, ribeiro2012sampling, cooper2014estimating, cooper2016fast}. Our work is related to the well-developed literature on sampling nodes uniformly at random using random walks \cite{bash2004approximately, chiericetti2016sampling}. However, in our work we sample entire $k$-edge walks uniformly at random from all walks on $k$-edges, which is a fundamentally different problem. The work that is most closely related is about sampling induced subgraphs or graphlet realizations on a specific number of vertices \cite{lu2012sampling, bhuiyan2012guise}. However, our work differs from these in that the subgraphs we are sampling are a subset of all possible subgraphs, since we only sample $k$-edge subgraphs that are valid walks through the graph. 

Our contribution is to link DeBruijn graphs as models of higher-order correlations in pathway data to counting and analyzing sequential motifs from trajectory data. In \cref{sec:method}, we describe our methodology for computing sequential motifs and evaluating their importance using DeBruijn graphs.

\section{Proposed Methods}
\label{sec:method}
\input{method}


\section{Experiments}
\label{sec:results}
\input{results}

\section{Conclusion}
\label{sec:conclusion}
We introduced a method, CAC, for counting and analyzing \emph{sequential motifs} on $k$ edges in observed walk data using DeBruijn graph ensembles and random walks. A major advantage of sequential motifs is that their interpretation is straightforward: we count the prevalence of a particular structure \emph{in the data}, not after aggregating it into a first-order network, thus throwing away sequential information. We presented three sampling procedures to assist in evaluating the importance of sequential motifs: (1) a uniform $k$-edge walk sampling procedure from a complete DeBruijn graph $G^k_c$, (2) a uniform sampling procedure over observed walks based on the edges of the observed DeBruijn graph $G^k$, and (3) a sampling procedure based on the Generalized Hypergeometric Ensemble of DeBruijn Graphs defined in \hypa~\cite{larock2020hypa}. We showed that sequential motifs are substantially different than static network motifs, and analyzed two datasets, highlighting the subtleties between the various null models and comparing to simply sampling random walks on the 1st-order graph.

There are numerous future directions for this research. For example, we can design bespoke DeBruijn graph null models for specific datasets based on external correlations such as geographical or conceptual distance between nodes. The convergence of our sampling methods to the true motif distribution, as well as the related question of when a random walk ensemble is good enough, also remain open. 

{\footnotesize
	\paragraph{Funding}
	IS acknowledges financial support by the Swiss National Science Foundation through grant 176938. TL and TER were funded by the Combat Capabilities Development Command Army Research Laboratory through Cooperative Agreement W911NF-13-2-0045 and by the Under Secretary of Defense for Research and Engineering under Air Force Contract No. FA8702-15-D-0001. Any opinions, findings, conclusions or recommendations contained in this document are those of the authors and should not be interpreted as representing the official policies, either expressed or implied, of the Combat Capabilities Development Command Army Research Laboratory, Under Secretary of Defense for Research and Engineering, or the U.S. Government. The U.S. Government is authorized to reproduce and distribute reprints for Government purposes not withstanding any copyright notation hereon.

	\paragraph{Acknowledgements}
	TL thanks Vahan Nanumyan for preliminary discussions and analyzes that led to this work; Leo Torres for advice and discussion about the methodology; and Brennan Klein for help with designing the visualizations.
}

\bibliographystyle{abbrv}
\bibliography{references}

\newpage
\appendix

\section{Parallel Construction Algorithm}
\label{app:parallel}
As mentioned in \Cref{sec:method}, the for loops in \cac\ can be parallelized. We note that constructing the DeBruijn graph from the list of sequences $S$ is the same as counting n-grams, a problem that has been extensively studied and parallelized \cite{pibiri2019handling}. In particular, we can do the counting within a Map-Reduce framework \cite{dean2008mapreduce}, where the mapping step splits each walk $w\in \mathcal{S}$ into length $k$ segments (edges in the $k$th order DeBruijn graph), while simultaneously projecting each segment into the alphabet $\Sigma$. Then the Reduce step aggregates the count of each segment (i.e., edges of the DeBruijn Graph) as well as each projected motif. \Cref{alg:mapreduce} shows the pseudocode for this procedure.

\begin{algorithm}
	\begin{algorithmic}[1]
		\Function{Map}{$\mathcal{S}=(w_1, w_2, \cdots, w_n)$}
		\For{ $w_i$ in $\mathcal{S}$}
		\State $\ell = $ length of $w_i$
		\LineComment{Split into length $k$ segments and project each}
		\For{$j$ in $0, \dots \ell-k+1$ }
		\State kseg $\gets$ $w_i[j, j+k]$ 
		\State $m \gets$ ProjectMotif(kseg)
		\State Reduce(kseg, 1)
		\State Reduce($m$, 1)
		\EndFor
		\EndFor
		\EndFunction
		\Statex
		\LineComment{Increment the count of $x$, which is either a sequence or a motif, by $f$}
		\Function{Reduce}{$x, f$}
		\State IncreaseCount($x, f$)
		\EndFunction
	\end{algorithmic}
	\caption{Map and Reduce functions for simultaneously constructing a DeBruijn graph and counting motifs.}
	\label{alg:mapreduce}
\end{algorithm}

\section{Recovery of Network Structures}
\label{app:structure}
\begin{figure}[h]
	\centering
	\includegraphics[width=\textwidth]{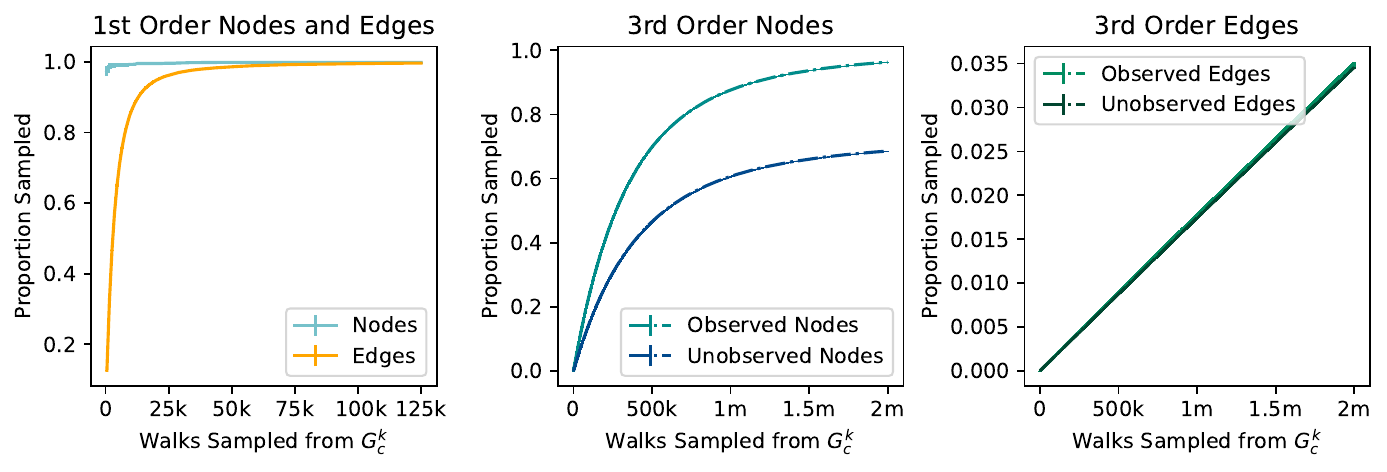}
	\includegraphics[width=\textwidth]{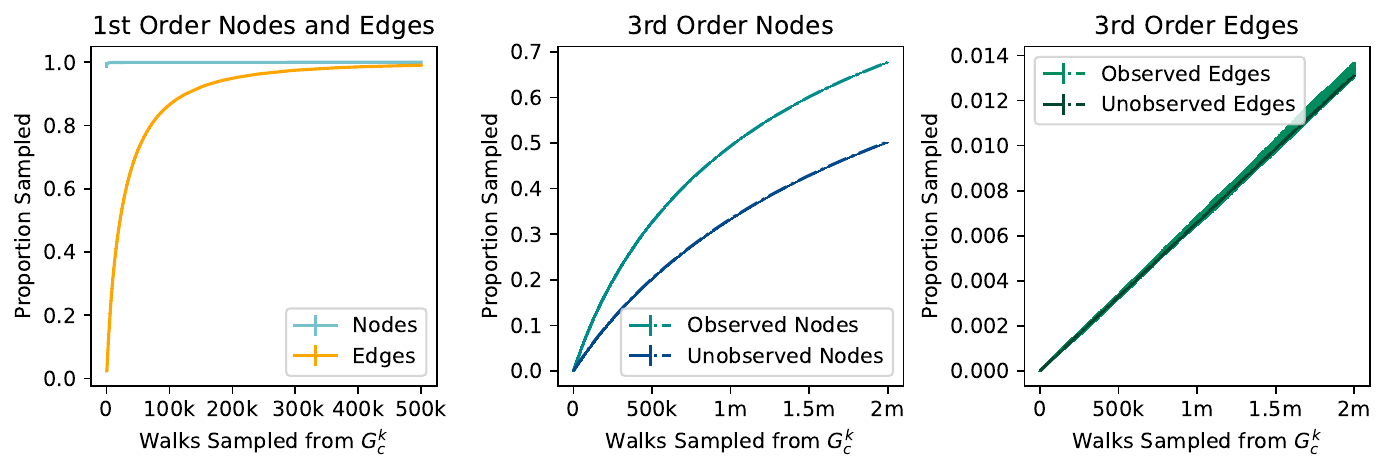}
	\caption{Relationships between the number of walks sampled uniformly from $G^k_c$ and recovery of network structure in the flights (top) and Wikispeedia (bottom) datasets. Left panel: first-order structure is recovered relatively quickly, after about 75k walks in the flights and 500k walks in Wikispeedia. Middle panel: the proportion of observed 3rd-order nodes appearing in sampled walks reaches 100\% in the flights dataset and around 70\% in the Wikispeedia after 2 million samples, while 60\% and 50\% of unobserved nodes appear in the same samples. Right panel: the number of observed and unobserved 3rd-order edges increases linearly with the number of walks sampled.}
	\label{motifs:fig:structure}
\end{figure}

In this section we briefly examine the relationship between sampling uniformly from $G^k_c$ and recovering network structure, both 1st and higher-order. In the first column of \Cref{motifs:fig:structure}, we show for each dataset the proportion of observed 1st-order nodes and edges recovered after sampling increasing number of walks from $G^k_c$. In both the flights (top) and Wikispeedia (bottom) datasets, all 1st-order nodes appear after a relatively small number of walks, while the whole edgeset is not sampled until about 75k walks have been sampled in the flights, and 500k walks sampled in Wikispeedia. 

In the middle column of \Cref{motifs:fig:structure}, we show the proportion of observed and unobserved 3rd-order nodes observed as the number of walks sampled from $G^k_c$ increases. Note that we analyze the 3rd-order rather than the 2nd-order network because unobserved nodes in the 2nd-order network would be unobserved edges in the 1st-order network, which will never appear by definition. After sampling 2 million walks uniformly at random, we discover all of the observed 3rd-order nodes in the flights dataset, but only about 60\% of the unobserved nodes. In the Wikispeedia data, after 2 million walks we have only sampled about 70\% of the observed nodes and 50\% of unobserved nodes.

Finally, in the last column of \Cref{motifs:fig:structure}, we show the same relationship as above but for 3rd-order edges. After 2 million samples, the proportion of sampled edges, both observed and unobserved, is less than 1\% for both datasets. The relationship between walks sampled and edges observed is linear, which is expected since walks are being sampled uniformly, and the number of walks sampled is much smaller than the total number of possible walks. This means that on average each sample adds 1 to either the observed or unobserved count.

\section{Approximation Procedures}
\label{app:approx}
We reduce the computational burden of sampling from $G^k_c$ by using the following procedure to sample edges uniformly without generating the entire DeBruijn graph. First, we sample a node $u$ from the set of all nodes in $V$ that are the source of at least one k-edge walk. Formally, we compute the number of walks originating from a node $u$ by taking the row summation of the $k$th power of the adjacency matrix of the directed graph $G$, which we will denote as $w_u = \sum_v \mathbf{A}^k_{uv}$. We then sample from the set of nodes where $w_u > 0$ with probability proportional to $w_u$, and generate all k-edge walks starting from $u$ by constructing all walks of length one--e.g., all of the directed edges $(u,v)$ pointing away from $u$--and iteratively expanding the set of walks by one edge--e.g., appending each neighbor of $v$ to the end of the previous walk--until the walks become longer than $k$-edges. Finally, we sample one of these walks uniformly at random. 

We argue that the above sampling procedure is equivalent to sampling from all edges in $G^k_c$. We note that the probability of sampling a given first-order walk from $G^k_c$ is simply the probability of choosing any of its $m$ edges: $p_{e_i} = \frac{1}{m}$. In contrast, our procedure requires two steps. First, we sample a starting node $u$ with probability $\frac{w_u}{\sum_{v\in V}w_v}$. Second, we sample one of the $w_u$ k-edge walks starting from $u$ uniformly. The probability of sampling an arbitrary walk $w$ starting from any node $u$ is the product of these two terms: 
$$p_w =  \frac{w_u}{\sum_{v\in V}{w_v} }\frac{1}{w_u} = \frac{1}{\sum_{v\in V}{w_v}} = \frac{1}{m} = p_{e_i}.$$
Thus the probability $p_w$ of sampling a k-edge walk $w$ using our two-step procedure is equal to the probability of sampling the same walk as an edge $e_i$ from the complete DeBruijn graph $G^k_c$. 

While the above procedure is more computationally tractable than constructing $G^k_c$ at the outset, if the number of walks we want to compute is substantially larger than $N$, we will implicitly construct $G^k_c$ anyway, since we will eventually compute all $k$-edge walks starting from all nodes $u$. To further reduce the computational burden, we use random walk simulations to sample. Instead of constructing \emph{all} $w_u$ $k$-edge walks starting from a node $u$, we simulate some number $x<w_u$ random walks, rejecting any walks with fewer than $k$ edges (e.g., if a walker arrives at a node with no out-degree). Then we sample a walk from this subset of random walks. If a node is sampled again, we compute another set of $x$ random walks, and union them with the previously computed set. Note that this procedure is no longer a uniform sample from the edges of $G^k_c$, since the probability of a random walk is determined by the out-degrees of its constituent nodes, meaning we are sampling a walk uniformly from a non-uniform sample of walks starting from $u$. We address this by sampling a random walk with probability proportional to the product of the out-degrees of the first $k$ nodes of the walk. This means that walks, which are individually less likely because they visit multiple hubs with many choices, are weighted more highly by our sampling procedure. This procedure will also eventually construct $G^k_c$ in its entirety, but we can now trade between speed and uniformity of the sample by setting $x$ lower (faster) or higher (more accurate) compared to the true number of walks per node.

An even simpler approximation is to simulate $M$ $k$-edge random walks on the first-order graph $G$, both with and without edge weights. This corresponds to comparing the observed DeBruijn graph to a version were all of the unobserved edges (including between nodes that were not observed in $G^k$) have weight proportional to their probability to be observed in $G$. This method also benefits from the familiarity of interpretation of random walks on graphs.

\subsection{Numerical Convergence in Synthetic Data}
\begin{figure}[!ht]
	\centering
	\includegraphics[width=\textwidth]{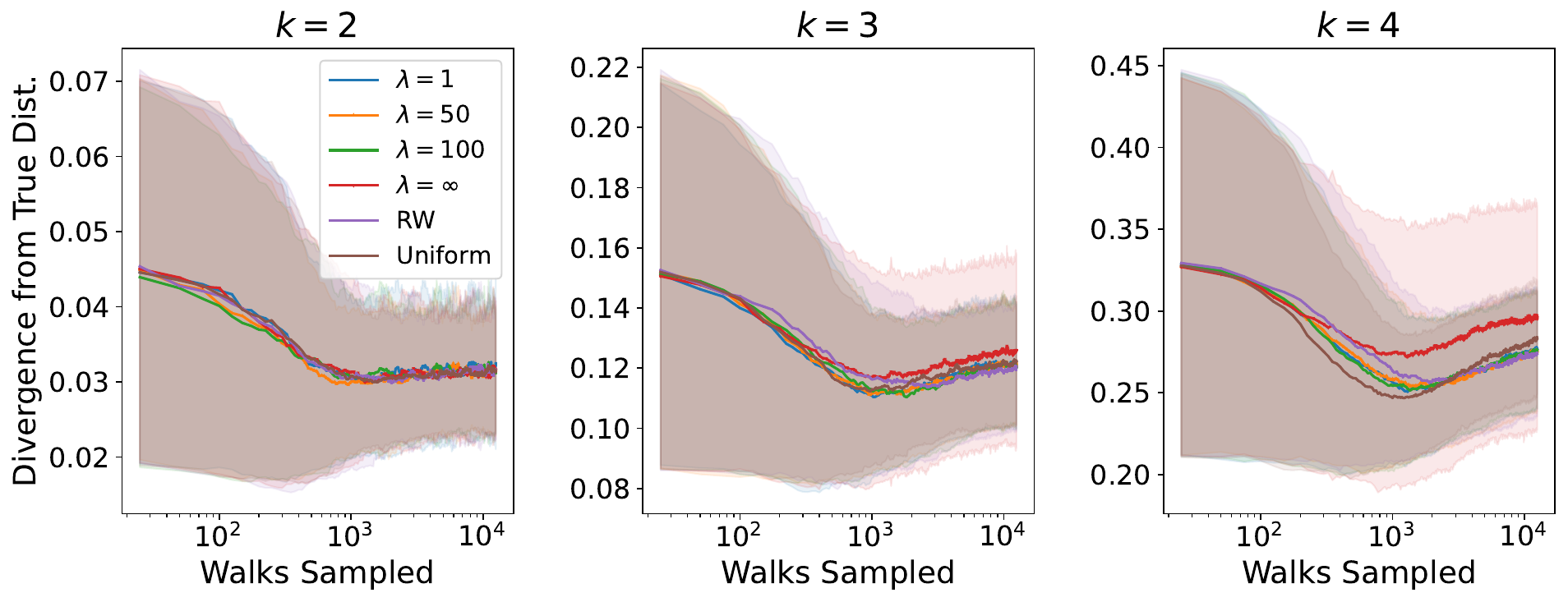}
	\caption{Convergence of sampled motif distribution towards the true motif distribution for orders $k=2,3,4$. The horizontal axis shows the cumulative number of walks sampled. The vertical axis shows the KL-divergence between the true motif distribution and the sampled motif distribution after accumulating the corresponding number of walks on the horizontal axis. The parameter $\lambda$ indicates the number of random walks computed per walk sampled for the approximate methods, with $\lambda=\infty$ corresponding to sampling one walk from all possible walks for each sampled node. ``RW" means sampling a single random walk in the 1st-order graph $G$ at every step. ``Uniform" means sampling a walk uniformly from all possible $k$-edge walks based on $G^k_c$. All methods converge to the uniform distribution at similar rates, but as $k$ increases, we see that the true uniform sampling method converges more quickly on average.}
	\label{motifs:fig:synthetic}
\end{figure}
We evaluate the efficacy of the sampling procedures described above in the following way. First, we generate a random network $G$ from $G_{N,p}$ \cite{gilbert1959random, erdos1960evolution}, with $N=500$ and $p=0.005$. For orders $k=2, 3, 4$, we compute the true motif distribution by computing the complete DeBruijn graph of $G$ and mapping all of its edges--all $k$-edge walks--to motifs, then normalize the motif counts to sum to 1. We then choose integers $w$ and $i$, where $i$ represents the number of iterations and $w$ represents the number of walks sampled per iteration. For every iteration we sample $w$ walks using each sampling method, map the walks to motifs, and accumulate the count of each motif for each method. Then we use the accumulated motif counts to compute a sample distribution and compute the KL-divergence between the sampled distribution up to iteration $j$ and the true distribution.

We show the results of this simulation using $w=25$ and $i=500$ in \Cref{motifs:fig:synthetic}, where each panel corresponds to an order $k$. In each plot of \Cref{motifs:fig:synthetic}, the horizontal axis represent the cumulative number of walks sampled so far--i.e., $w, w\cdot2, \cdots, w\cdot i$--and the vertical axis represents the KL-divergence between the true motif distribution and the sampled distribution so far, averaged over 1000 samplings. Each line corresponds to a different sampling procedure: $\lambda=\infty$ refers to computing all $k$-edge walks for every sampled start node; $\lambda=1, 10, 50, 100$ refer to computing $\lambda$ walks per sampled node; ``RW" refers to sampling a single random walk from the 1st-order graph $G$; and ``Uniform" refers to sampling one walk from all possible $k$-edge walks based on $G^k_c$. At $k=2$, where there are only 2 sequential motifs, the KL-divergence is already close to the true distribution after only a small number of walks are sampled. However, for $k=4$, we see that the uniform sampling has a slightly lower average for the first 1000 samples.

\section{Additional Results: Wikispeedia Unfinished and Flights Q2}
\label{sec:appC}
\begin{figure}[ht]
	\centering
	\includegraphics[width=0.55\textwidth]{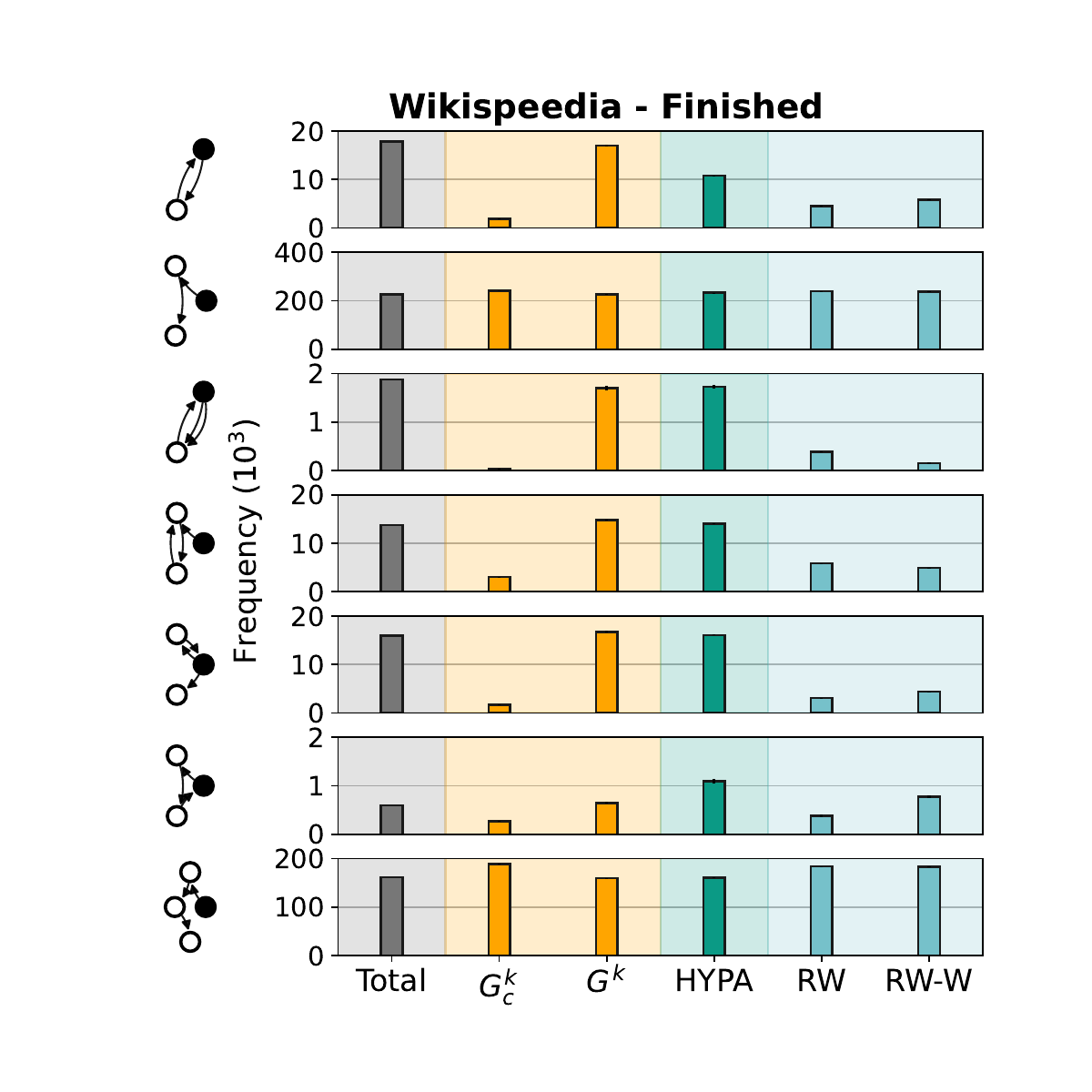}
	\hspace{-2.1cm}\includegraphics[width=0.55\textwidth]{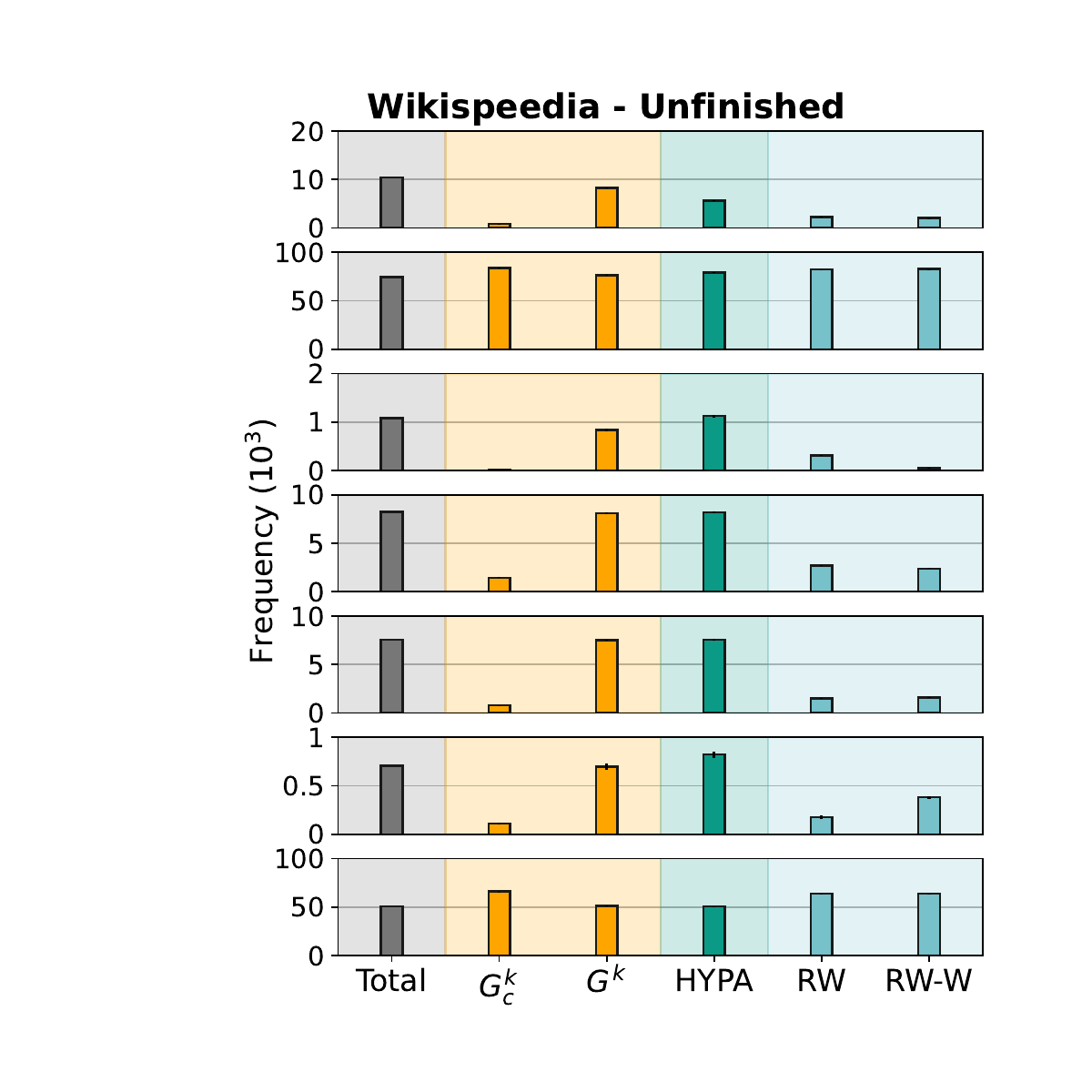}
	\caption{Sequential motif results for Finished Wikispeedia games (left, reproduced from \Cref{fig:motifs}) and Unfinished games (right). Directed triangles (row 6 of the figure) play different roles in finished and unfinished games.}
	\label{app:wiki}
\end{figure}

In \Cref{app:wiki}, we reproduce the Wikispeedia - Finished results from \Cref{fig:motifs} on the left, and show the results for Wikispeedia - Unfinished on the right. The results are very similar overall, however there is 1 key difference worth highlighting. The directed triangle motif (6th row of \Cref{app:wiki} appears at similar frequencies in both finished and unfinished games (around 500 observations), but in the finished games, the uniform ensemble suggests that the motif is either within expectation or slightly under-represented, while in the unfinished games the motif appears over-represented. This could indicate that triangles are indicative of a bad game, because the player makes two steps, gets frustrated with their options, then returns to an earlier node.

\begin{figure}[ht]
	\centering
	\includegraphics[width=0.55\textwidth]{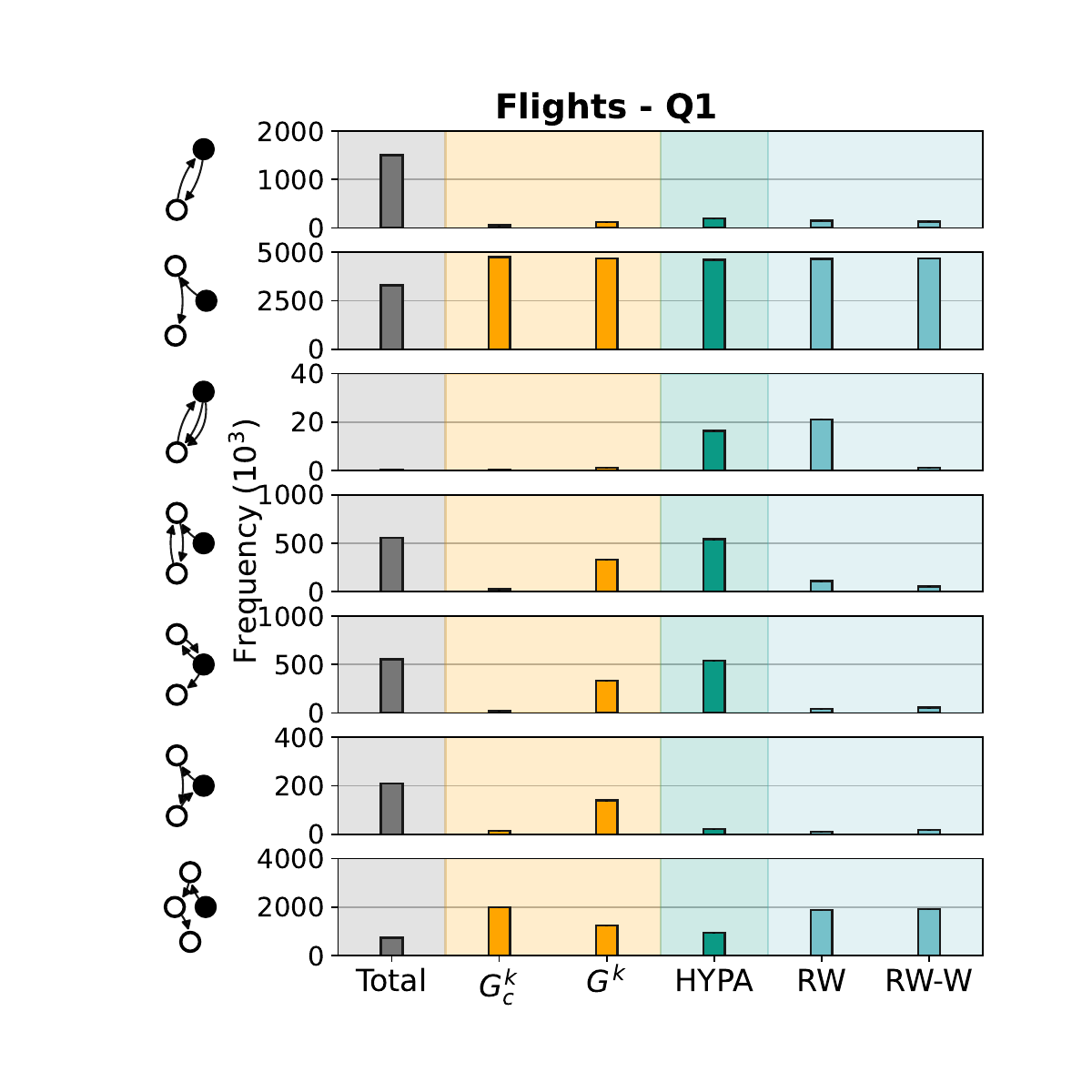}
	\hspace{-2.1cm}\includegraphics[width=0.55\textwidth]{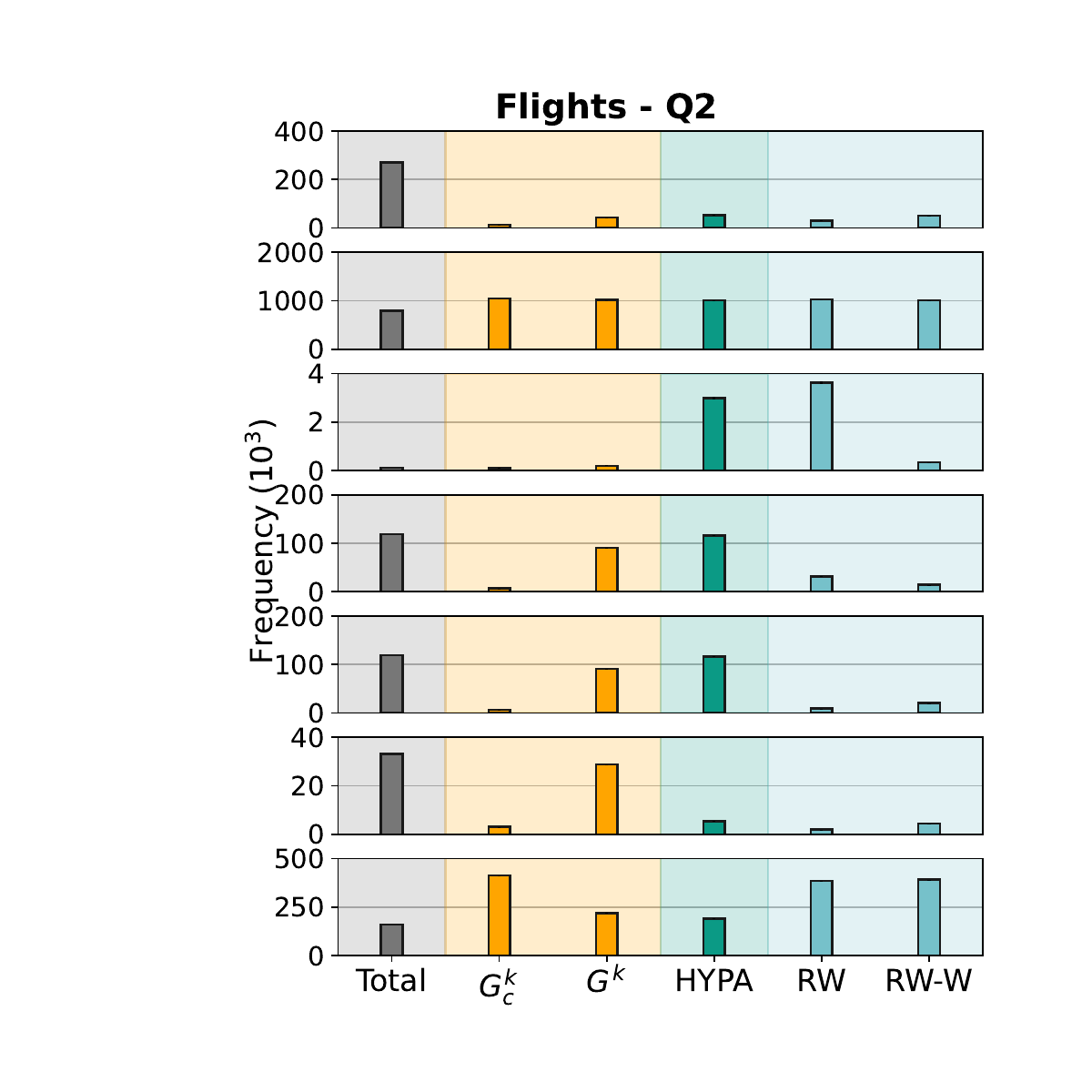}
	\caption{Sequential motif results for flights in Quarter 1 of 2020 (left, reproduced from \Cref{fig:motifs}) compared to Quarter 2 (right). Comparing Q1 and Q2, we observe chain motifs (row 2) in Q2 at rates much closer to the expectation the null models. This might indicate a preference on the part of passengers for 1-way trips to destinations where they planned to stay during the pandemic, rather than round trips that are typical according to Q1 data.}
	\label{app:flights}
\end{figure}

In \Cref{app:flights}, we compare flight motifs between Q1 (left, reproduced from \Cref{fig:motifs}) and Q2 (right) of 2020. The overall number of flights dropped considerably, likely due to the COVID-19 pandemic. Specifically, the use of the 2-node chain motif (2nd row in \Cref{app:flights}) decreased both in terms of observed trips and in the extent to which it is underrepresented, potentially indicating that passengers were opting for 1-way trips during the pandemic. Future work should examine these patterns while controlling for year-to-year seasonality.

\end{document}

%% file: method.tex
In this section we first present our method for simultaneously constructing a $k$th order DeBruijn graph and counting $k$-edge sequential motifs, which we call \textsc{ConstructAndCount}, or \cac. Then we describe a method for evaluating the importance of motifs using randomizations of complete DeBruijn graphs.

\subsection{Sequential Motif Counting}
We start from a dataset of $n$ trajectories $\mathcal{S} = \{(s_1, w_1), (s_2, w_2), \cdots (s_n, w_n)\}$ through the directed graph $G=(V,E)$ defined by the edges in $\mathcal{S}$. We are also given an alphabet $\Sigma=(\sigma_1,\sigma_2,\dots,\sigma_{\ell})$ of arbitrary symbols (see problem statement in \cref{sec:intro} for more details). Finally, we assume we have chosen an integer $k$ that will be the length in edges of sequential motifs we are interested in computing.

We provide the pseudocode for \cac\ in \Cref{alg:motifs}. \cac\ takes as input a dataset of sequences $\mathcal{S}$, motif alphabet $\Sigma$, and order $k$, and proceeds as follows. First, \cac\ initializes the set of $k$th-order nodes $V^k$ and edges $E^k$ to be empty. These will define the DeBruijn graph. \cac\ also initializes all entries of the lookup table $\mathcal{C}[m]$, indexed by motifs $m\in\mathcal{M}^k$, to 0. This lookup table  will contain the frequency counts of each motif. \cac\ then enters the outer for loop (line 3) that iterates over all pairs of sequences and their associated weights: $(s,w) \in\mathcal{S}$. The default weight is 1. In the inner for loop (line 4), \cac\ slides a window of length $k$ across $s$, where each index $i$ corresponds to the first position of a $k$-edge subsequence \emph{seq} starting at $s[i]$, computed in line 5. \cac\ then determines which motif the sequence \emph{seq} corresponds to by projecting it into the alphabet $\Sigma$ using the ProjectMotif(seq, $\Sigma$) procedure, which maps each node $u\in s$ to a distinct element $\sigma_j\in\Sigma$ in order of appearance, as illustrated in \cref{fig:example}. After a node has been mapped to a symbol, it is replaced everywhere in the projected path with that symbol. Once every node in $\textrm{seq}$ has been mapped, the procedure returns the sequential motif $m\in\mathcal{M}^k$ corresponding to \emph{seq}. Now \cac\ increments the count of $m$ in $\mathcal{C}$ by the frequency of $s$, constructs the edge $e$ from \emph{seq}, and adds the node, edge and frequency to $V^k$, $E^k$, and $W^k$. Note that paths observed in $S$ with length shorter than $k$ can not be included by definition, since $|s|-k$ will be negative and thus the inner for loop will not begin. 

\begin{algorithm}
	\caption{\textsc{ConstructAndCount}($S,k,\Sigma$): Construct the $k$th order weighted DeBruijn Graph \& count all length $k$ motifs from sequence dataset $S$ using motif alphabet $\Sigma$.}
	\label{alg:motifs}
	\begin{algorithmic}[1]
		\Require $S = \{(p_1,w_1), (p_2,w_2)\cdots,(p_N,w_N)\}$ (sequence dataset), $k$ (order)
		\Ensure $G^k \gets (V^k, E^k, W^k)$ (DeBruijn graph), $\mathcal{C}[m]$ (motif counts)
		\State $V^k, E^k\gets \emptyset$
		\State $\mathcal{C}[m] \gets \emptyset$ $\forall_{m \in \mathcal{M}^k}$
		\For{walk $s\in S$ with weight $w_s$}
			\For{$i$ from $0,\cdots,|s|-k$}
				\State seq = $s[i, \cdots,i+k]$
				\State $m \gets$ ProjectMotif(seq, $\Sigma$)
				\State $\mathcal{C}[m] \gets \mathcal{C}[m] + w_s$
				\State $\vv{u} \gets \textrm{seq}[0,\cdots, k-1], \vv{v} \gets \textrm{seq}[1,\cdots,k]$
				\State $V^k \gets V^k \cup \{\vv{u}, \vv{v}\}$; $E^k \gets E^k \cup (\vv{u}\vv{v})$
				\State $W^k_{\vv{u}\vv{v}} = W^k_{\vv{u}\vv{v}} + w_s$
			\EndFor
		\EndFor
		\\
		\Return $G^k=(V^k, E^k, W^k), \mathcal{C}$
	\end{algorithmic}
\end{algorithm}

The runtime of \cac\ depends on three variables: (i) $n$, the number of sequences in $\mathcal{S}$; (ii) $p(\ell)$, the distribution of lengths of sequences in $\mathcal{S}$; and (iii) the motif length $k$. Assuming all paths are of the maximum length $\ell_{\max}$, the worst case time is $O(n\cdot (\ell_{\max}-k))$. We note that in practice distributions of path lengths tend to have tails in large values, meaning that in real data the average path length $\ell_{\textrm{avg}}$ is likely to be much lower than the maximum (see Figure~\ref{fig:pathlengths}). Lastly, we note that the two loops in the \cac\ algorithm can also be parallelized over the edges, since the operations for each edge (lines 4 through 10) are independent of any other edge (see \Cref{app:parallel}).

\subsection{Measuring Motif Importance via Random DeBruijn Graphs}
\label{subsec:importance}
To evaluate the importance of a motif, we compare the count of that motif in the observed data with the average count of the same motif in many random samples from a null model. This is a flexible framework for determining importance, since the null model we choose determines which properties of the input data are randomized and to what extent, and therefore null models can be designed to test different hypotheses about the processes generating the observed data (similar to \cite{milo2002network, artzy-randrup2004comment}). Here we propose three null models. First, we describe a null model that relies on sampling from a \emph{complete $k$th-order DeBruijn graph $G^k_c$}, where all possible higher-order edges based on the directed first-order topology $G=(V,E)$ can be sampled with non-zero probability. Second, we propose a null model that samples uniformly from all \emph{observed} $k$-edge walks by sampling from the edges of the observed DeBruuijn graph $G^k$. Finally, we adopt the null model proposed in \cite{larock2020hypa}, which samples from the edges of the observed $k$th-order DeBruijn graph based on a $k-1$st-order null model.

\subsubsection{Sampling Uniformly at Random from All Possible $k$-edge Walks}
\begin{figure}[h]
	\centering\includegraphics[scale=0.85]{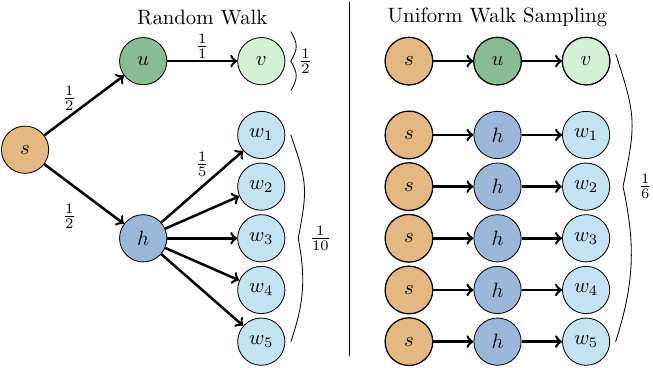}
	\caption{Sampling $2$-edge random walks over the graph on the left is not the same as sampling uniformly from all possible $2$-edge walks.}
	\label{fig:rw-motivation}
\end{figure}

Sampling from $G^k_c$ is advantageous because it is conceptually simple to construct a complete $k$th-order DeBruijn graph $G^k_c$ from a directed graph $G$ that includes every possible $k$th order node and edge. The procedure works as follows: first, take all of the edges in $G$ and turn them into nodes in $G^2_c$; then, add edges between higher-order nodes $(s,u)$ and $(v,t)$ if $u=v$; and finally, repeat this procedure $k$ times.\footnote{For further intuition, we note that the complete $k$th-order DeBruijn graph is equivalent to the line graph transformation applied $k$ times to the 1st-order graph $G$.} Sampling edges from this graph is equivalent to sampling a $k$-edge walk uniformly at random from all possible $k$-edge walks through $G$.\footnote{Note that our goal is \emph{not} to sample nodes uniformly at random, which is a well-studied problem. In fact, we do not expect nodes to be sampled uniformly in this setting, since nodes with higher degree will be involved in more walks.} Importantly, this procedure is \emph{not} equivalent to computing random walks on the network topology. In \Cref{fig:rw-motivation} we show a concrete case where sampling $k$-edge random walks would not results in a uniform distribution over all possible walks.

In practice, computing complete DeBruijn graphs $G^k_c$ for even relatively small values of $k$ is computationally expensive, since the number of nodes and edges grows exponentially. In \Cref{app:approx}, we give two strategies for sampling approximately uniformly from the edges of $G^k_c$.

\subsubsection{Sampling Uniformly at Random from Observed $k$-edge Walks}
Sampling from all possible $k$-edge walks using $G^k_c$ is maximally random while maintaining the 1st-order network structure. This is both an advantage, because it gives us a sense for what we could expect if the observed motifs were based on fully random data, but also a disadvantage, since many of the walks sampled were not observed in the real data. As a first step towards a more data-informed null model, we also sample uniformly at random from the edges of the observed DeBruijn graph $G^k$. This is a uniform random sample over the observed $k$-edge walks. If all of the edge weights in $G^k$ are equal to 1, then our sampled motif distribution should be approximately equal to the observed motif distribution. However, since this is not the case in our datastets, sampled counts for motifs corresponding to edges in $G^k$ that have high weights are likely to be substantially smaller, and vice-versa.

\subsubsection{Sampling $k$-edge Walks from \hypa}
The previous two null model sought to sample uniformly from a set of $k$-edge walks. We also evaluate motif importance by sampling from ensemble derived from previous work on anomaly detection in DeBruijn graphs, which corresponds to sampling $k$-edge walks from a $k-1$st-order model of the observed data \cite{larock2020hypa}. \hypa\ represents a soft configuration model for the edge weights of a $k$th order observed DeBruijn graph, called the Generalized Hypergeometric Ensemble of Random Graphs \cite{casiraghi2017relational}. Samples from this ensemble preserve the average in and out weight for each node.

We note that each of the null models we have proposed here can be customized to incorporate known correlations. For example, we may observe spatial correlations that indicate a high likelihood of certain walks occurring, and so rather than sampling uniformly, we may weight the walks based on a measure that captures this correlation. In fact, the Generalized Hypergeometric Ensemble that underlies \hypa\ is designed to incorporate this information using the propensity matrix \cite{casiraghi2017relational}. We leave investigation of these bespoke null models for future work.

%% file: results.tex
In this section, we present experiments on counting and analyzing sequential motifs in two datasets. We first describe the datasets--a transportation network and an online hyperlink navigation network--then we compare sequential motifs within and across the datasets. Appendix~\ref{sec:appC} contains more experimental results, and an implementation of our methodology is available online \cite{larock2021debruijn}.

\subsection{Data Description}
Our first dataset consists of a large sample of flight itineraries representing trajectories of passengers through the domestic airport network in the United States in Quarter 1 of 2020 \cite{transstat2018origin}. Each itinerary is a walk through the network corresponding to a starting airport, $i\geq0$ layover airports, and a destination. We are also given frequencies for each walk corresponding to the number of people who bought a ticket with that itinerary (e.g., a number $w$ of people bought tickets with the itinerary JFK to Chicago O'Hare to Seattle, Washington). 

\begin{figure}[h]
	\centering\includegraphics[scale=0.75]{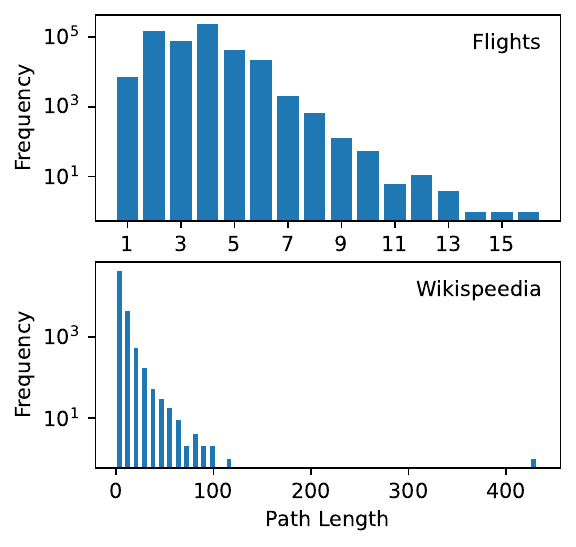}
	\caption{Histograms of sequence lengths in the passenger flights (top) and Wikispeedia (bottom) datasets. Both datasets have heavy tails in longer sequences, with most observations being relatively short and a small number of very long walks. Walks through the airport network are naturally shorter with a maximum of 10, while Wikispeedia walks have more variable length, with a relatively small number taking over 100 steps before finding the target page.}
	\label{fig:pathlengths}
\end{figure}

The second dataset consists of walks through the Wikipedia network during successful runs of the Wikispeedia game, where a player was given a random target page in Wikipedia, then placed on a random start page and asked to navigate from the start to the target using only internal Wikipedia links \cite{west2012human}. These walks are not associated with frequencies, though we note that the DeBruijn graph edges are still weighted, since the same paths are repeated in different instances of the game. We present statistics of each of the datasets in \Cref{tab:seqstats}, and show the distribution of sequence lengths in each dataset in \Cref{fig:pathlengths}.

\begin{table}[h]
	\centering
	\caption{Statistics of sequence datasets. Recall that $|V|$ is the number of nodes and $|E|$ is the number edges. $l_\textrm{avg}$ and $l_\textrm{max}$ are the average and maximum length of paths, respectively.  }
	\label{tab:seqstats}
	\begin{tabular}{|l|l|l|l|l|l|}
		\hline
		Dataset & $|\mathcal{S}|$ & $\ell_{\textrm{avg}}$ & $\ell_{\max}$ & $|V|$ & $|E|$ \\ \hline
		Flights & 1,789,020 & 3.4 & 16 & 433 & 10,954 \\ \hline
		Wikispeedia & 282,863 & 5.9 & 434 & 4,169 & 59,530 \\ \hline
	\end{tabular}
\end{table}

\subsection{Sequential Motifs in Flight Patterns and Online Navigation}

\begin{figure*}[!ht]
	\includegraphics[width=0.53\textwidth]{{figures/coupons_2020_q1.ngram}.pdf}\hspace{-2.0cm}
	\includegraphics[width=0.53\textwidth]{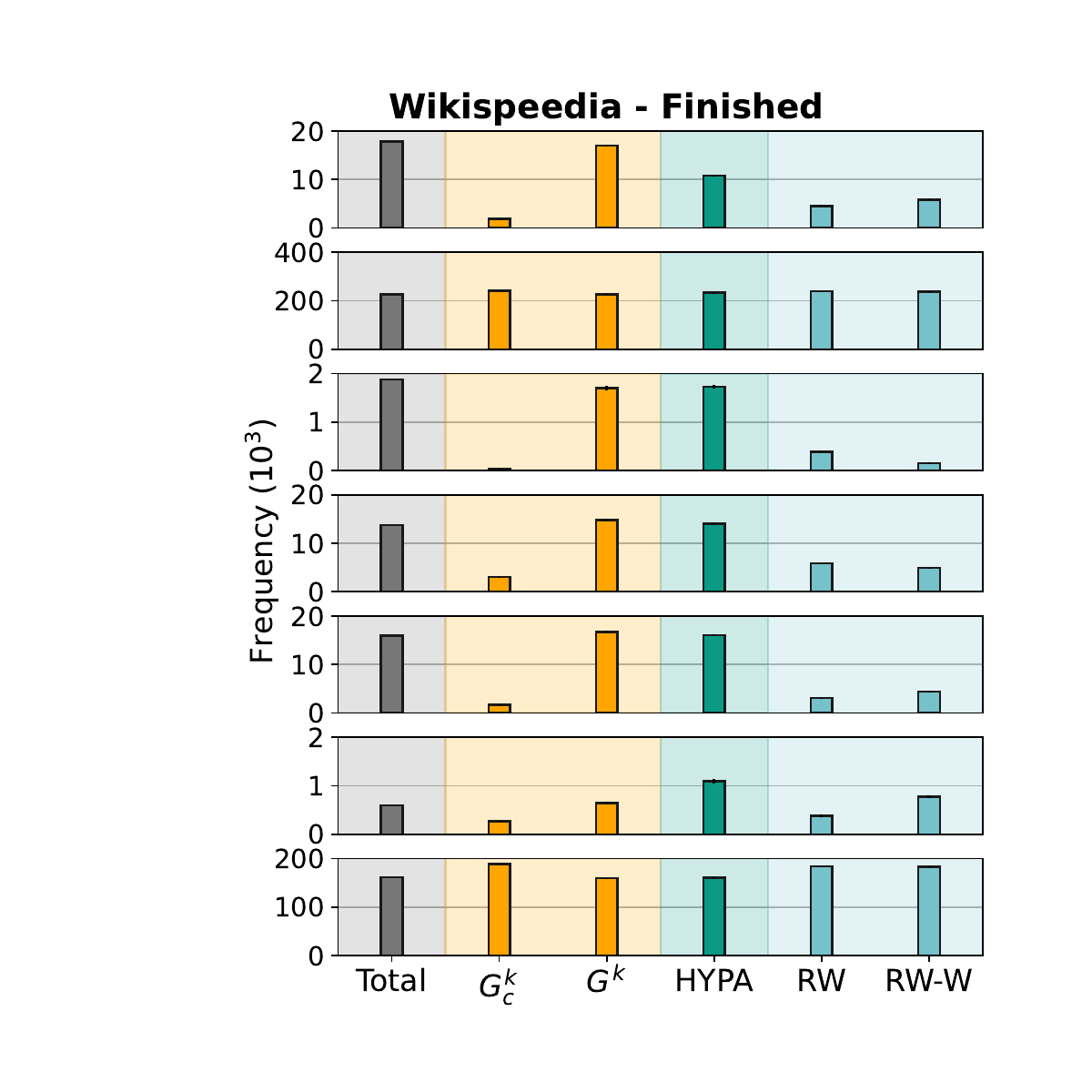}
	\caption{Comparing motif importance in US domestic flight paths (left) and paths through the Wikipedia page network in successful Wikispeedia games (right). Each row corresponds to a sequential motif (shown on the left). The gray panel shows the total number of motifs counted in the data. Note that the y-axes are measured in thousands. The orange bars show the average counts when sampling uniformly from the $G_c^k$ and $G^k$ ensembles; the dark green bar shows the average counts when sampling from \hypa\; and the blue bars show counts from unweighted (RW) and weighted (RW-W) random walks. A sequential motif is \emph{overrepresented} with respect to a null model if its observed frequency is greater than its frequency in the null model, and \emph{underrepresented} if its observed frequency is less than its frequency in the null model. In the Flights data, directed triangles are over-represented according to all ensembles. However, the results are mixed for the same motif in the Wikispeedia data, where it is considerably less prevalent. All null models except for the \hypa\ ensemble show the directed triangle as either within expectation or slightly underrepresented, while the \hypa\ ensemble suggests that the triangle is very underrepresented, appearing about twice as much in random samples. Since the directed triangle is a 3-edge sequential motif, the corresponding \hypa\ ensemble preserves the frequency of 2nd-order patterns in the Wikispeedia data. The fact that more triangles would be expected under this null model means that players opt not to close triangles as often as we would expect at random.}
	\label{fig:motifs}
\end{figure*}

In \Cref{fig:motifs}, we show the distribution of sequential motifs for orders 2 and 3 in both of our datasets. Each row of the figure corresponds to a single motif. The gray bar shows the observed frequency of each motif. The remaining bars show the average count of the motif over 10 randomized datasets from each of the following ensembles: uniformly from $G^k_c$; uniformly from $G^k$; the \hypa\ ensemble; unweighted first-order random walks; and weighted first-order random walks.\footnote{We also computed standard deviations from the averages and found that they are typically two or more orders of magnitude smaller than the averages, making error bars too small to show on the figures.} We will consider a motif \emph{overrepresented} with respect to a null model if its observed frequency is greater than its frequency in samples from the null model, and \emph{underrepresented} if its observed frequency is less than its frequency in the null model. If a motif has similar frequency in both the observed data and null model samples, we say its frequency is within expectation.

We begin with some general observations about the motif distributions. First, we note that all sampling methods are biased towards chain motifs (second and last rows in \Cref{fig:motifs}). This is not surprising since the number of possible chain motifs on $k$ edges grows with the in- and out-degrees of the hubs, which are present in both networks.

A second observation is that sampling random walks is generally a good proxy for sampling uniformly from all possible walks. There are only a small number of motifs measured from our datasets where the unweighted random walks result in qualitatively different results (Flights, 3rd row and Wikispeedia, 6th row in \Cref{fig:motifs}). 

\begin{figure}[h]
    \includegraphics[scale=0.55]{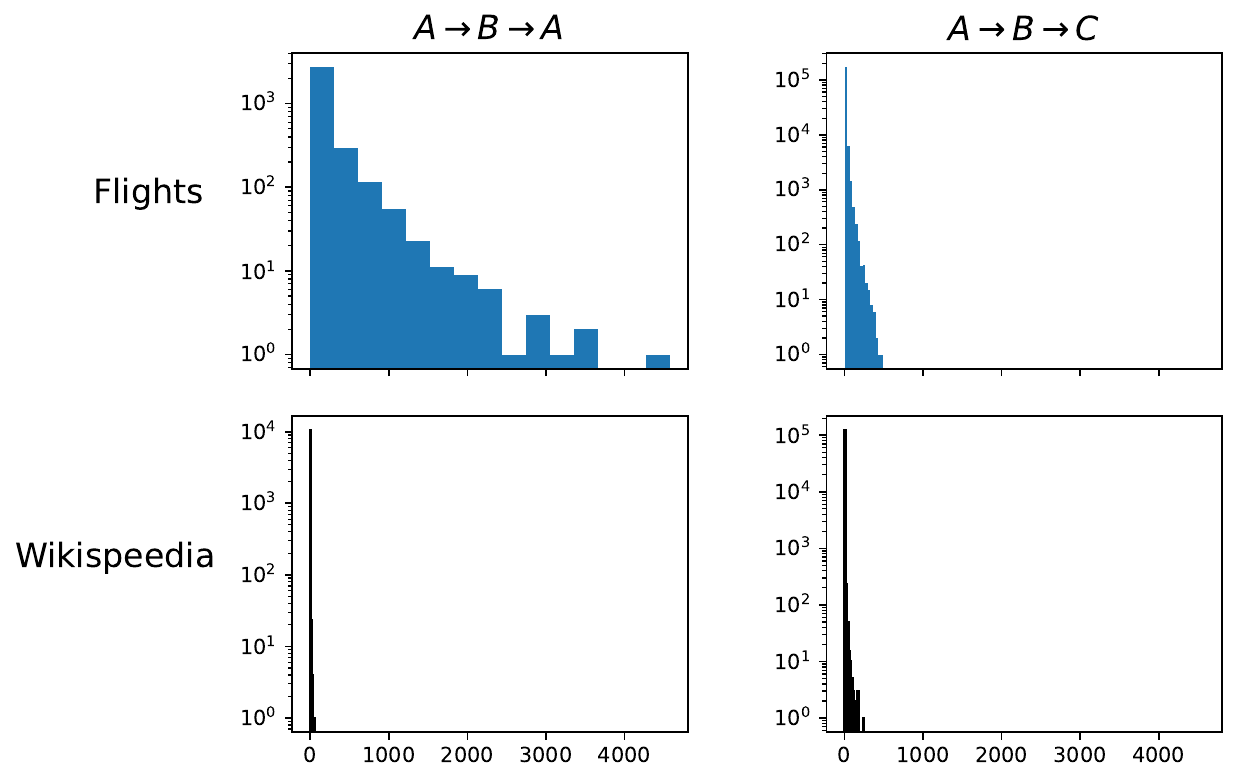}
	\caption{Histograms of weights for edges corresponding to the 2nd-order motifs in Flights (top) and Wikispeedia (bottom) datasets. The distribution of edge weights for the backtrack motifs in the flights data (top left, A-B-A) is heavy tailed, while all others are not. This explains why all null models result in similar predictions for the flights backtrack motif.}
	\label{fig:secondorder}
\end{figure}

We now compare the distributions of each motif in turn, beginning from the first 2-edge motif at the top of \Cref{fig:motifs}, which corresponds to the sequence A-B-A. In the flights data, paths that correspond to this motif represent round trips starting and ending at the same airport, whereas in the Wikispeedia game these correspond to a sort of ``guess and check" behavior, where a player starts at page A, chooses a promising page B, then finds that the page does not contain any links that are better than those on the previous page, and so clicks a link back to page A. The motif is prevalent in both datasets, and is also overrepresented based on almost all ensembles. In the flights data, overrepresentation indicates that round trips occur substantially more often than expected in the real data than in any of the null models. This is not surprising, since people presumably prefer to take fewer flights, thus a direct flight in both directions is ideal. However, it is worth noting that the results for 2-edge motifs in the flights data are essentially the same across all of the null models. To understand this, we first note that at $k=2$, all of the null models, including \hypa, are sampling some variation of a $1$st-order random walk. In fact, at $k=2$ the \hypa\ ensemble and the weighted random walk (RW-W) are sampling from the same distribution--random walks on the weighted 1st-order topology--but with different strategies. In the flights data, regardless of which null model we choose, the result is the same: backtracks are very overrepresented, and chains are very underrepresented. One implication of this pattern is that the weights on the 2nd-order edges that correspond to backtracks must be non-uniform in the flights data, otherwise we would expect samples from $G^k$ and \hypa\ to contain more backtracks. This makes intuitive sense because some round-trip flights--such as flights from major cities to Washington, D.C.--are likely to have extreme edge weights relative to less traveled trips. However, this does not hold in the Wikispeedia data, where sampling edges uniformly from the observed $G^k$ results in an average count of backtrack motifs that is close to the observed count, and sampling from \hypa\ results in considerably more backtracks than the other null models. This indicates that the weight of the 2nd-order edges corresponding to the backtrack motif in the Wikispeedia data are relatively small on average, since sampling uniformly from those edges results in approximately the same number of backtracks. Since the Wikispeedia network is larger and more sparse than the flights network, it is not surprising that the distribution of 2nd-order backtrack weights is more uniform, as we show in the bottom left panel of \Cref{fig:secondorder}.
 
The second 2-edge motif is the chain on 3 nodes: A-B-C. This motif is also prevalent in both datasets. In the flights data, the 2-edge chain appears underrepresented compared to all of the ensembles. As we observed previously, all of the random sampling procedures are biased towards chains, since there is a disproprtionate number of possible chains due to the presence of hubs. The relatively low prevalence in the flights data is the flip side of the observation we made above: since people prefer to take direct flights in both directions, chains occur less often than expected at random. In the Wikispeedia data, the motif appears at approximately the same rates as all of the ensembles predict. In contrast to the flights, where the goal is often to return to the origin, in the Wikispeedia network the goal is to move away from the source and towards the target, so while backtracks are overrepresented, 2-edge chains are also an important part of the searching process.

The third motif from the top is the first 3-edge motif, corresponding to a round trip that doubles back again (e.g., the sequence A-B-A-B). This motif is rarely observed in the flights dataset, and is observed at approximately the same rates as in samples from the $G_c^k$, $G^k$, and weighted random walk ensembles. However, the motif appears under-represented based on the \hypa\ and unweighted random walk ensembles. This is likely related to the prevalence of round trips in the observed data, which make the second backtracking edge much more likely than expected at random in a $2$nd-order model or weighted random walk. In the Wikispeedia data the motif appears to be within expectation for the observed $G^k$ and \hypa\ ensembles, but is over-represented with respect to the others. As we discussed previously, the $G^k$ and \hypa\ ensembles are constrained to walks that were observed in the real data, and so even the relatively small prevalence of this motif creates a substantial bias in samples from those ensembles.

We will analyze the next two motifs (rows 4 and 5 from the top in \Cref{fig:motifs}) together, since they are complementary to one another. The first represents the sequence A-B-A-C, while the second represents A-B-C-B. Taken together these motifs have an intuitive interpretation in the flights data: they represent a round trip with a layover at the same airport in each direction. In the first direction, we move from the source airport $a_1$ to a layover airport $a_2$, then on to our destination to $a_3$. On the return trip, we go from airport $a_3$ back to the layover airport $a_2$, before finally returning again to $a_1$. The 4th motif in \Cref{fig:motifs} corresponds to the first direction plus the edge $(a_3,a_2)$ going back to the layover airport, while the 5th motif corresponds to the same edge plus the rest of the trip. These motifs are overrepresented with respect to the $G^k_c$ and random walk ensembles in the flights data, which is intuitive since, as we noted in our discussion of the round trip motif, people tend to take round trips when they fly, plus the fact that airlines send flights between the same airports on a regular schedule, making it likely that the layover will be the same in both directions. Again we see that the ensembles constrained to the observed walks result in counts similar to the observed frequencies. However, in this case the uniform samples from $G^k$ still result in fewer observations of these motifs, while the counts based on \hypa\ are approximately the same as the observed data. The interpretation of these motifs in the Wikispeedia data is analogous if less intuitive, since they represent mediated round trips from one page, through two intermediary pages, then back where it started. This is indicative of a player giving up on a path and returning to an earlier page to try again, perhaps remembering that there was another promising link a few pages back. Accordingly, these motifs are also overrepresnted in the Wikispeedia sequences with respect to the $G^k_c$ and random walk ensembles, and just like in the flights data, the $G^k$ and \hypa\ ensembles produce counts much closer to the observations.

Next we analyze the directed triangle (row 6), compared earlier to its static counterpart in \Cref{fig:static_vs_seq}. In the flights data, this motif represents a trip with a direct flight in one direction and a layover in another, or a multi-city trip starting and ending at the same airport. In Wikispeedia, it corresponds to essentially the same discussion as in the last paragraph, but where the player did not need to backtrack because they found a link to the page to which they wanted to return. This motif is overrepresented with respect to all ensembles in the flights data, although the average count based on sampling uniformly from $G^k$ is greater than the rest, suggesting that observed $k$th-order edges that correspond to triangles are a larger share of the edges in $G^k$ compared to $G^k_c$. In the Wikispeedia data, where this motif occurs relatively rarely, all but the \hypa\ ensemble averages suggest the motif appears approximately within expectation, while the \hypa\ ensemble suggests the motif is underrepresented, indicating that the directed triangle is more likely to occur in $k$-edge walks through a $2$nd-order model of the data.

Finally, we come to the 3-edge chain on 4 nodes (row 7). This is a very common motif in both datasets, and it appears to be underrepresnted or within expectation according to all of the ensembles.

Taken together with \Cref{fig:static_vs_seq}, this analysis shows how sequential motifs can be used to understand a dataset in a way that is consistent with the inherent sequential nature of the process generating the data.

%% file: ms.bbl
\begin{thebibliography}{10}

\bibitem{abello2010detecting}
J.~Abello, T.~{Eliassi-Rad}, and N.~Devanur.
\newblock Detecting {{Novel Discrepancies}} in {{Communication Networks}}.
\newblock In {\em Proceedings of the 10th {{IEEE International Conference}} on
  {{Data Mining}}}, ICDM'10, pages 8--17, 2010.

\bibitem{ahmed2014network}
N.~K. Ahmed, J.~Neville, and R.~Kompella.
\newblock Network {{Sampling}}: {{From Static}} to {{Streaming Graphs}}.
\newblock {\em ACM Transactions on Knowledge Discovery from Data},
  8(2):7:1--7:56, 2014.

\bibitem{alon2007network}
U.~Alon.
\newblock Network motifs: Theory and experimental approaches.
\newblock {\em Nature Reviews Genetics}, 8:450, June 2007.

\bibitem{artzy-randrup2004comment}
Y.~{Artzy-Randrup}, S.~J. Fleishman, N.~{Ben-Tal}, and L.~Stone.
\newblock Comment on ``{{Network Motifs}}: {{Simple Building Blocks}} of
  {{Complex Networks}}" and ``{{Superfamilies}} of {{Evolved}} and {{Designed
  Networks}}".
\newblock {\em Science}, 305(5687):1107--1107, Aug. 2004.

\bibitem{bash2004approximately}
B.~A. Bash, J.~W. Byers, and J.~Considine.
\newblock Approximately uniform random sampling in sensor networks.
\newblock In {\em Proceedings of the 1st {{International Workshop}} on {{Data
  Management}} for {{Sensor Networks}} (Held in Conjunction with {{VLDB}}
  2004)}, pages 32--39, 2004.

\bibitem{battiston2020networks}
F.~Battiston, G.~Cencetti, I.~Iacopini, V.~Latora, M.~Lucas, A.~Patania, J.-G.
  Young, and G.~Petri.
\newblock Networks beyond pairwise interactions: {{Structure}} and dynamics.
\newblock {\em Physics Reports}, 874:1--92, 2020.

\bibitem{benson2019three}
A.~R. Benson.
\newblock Three {{Hypergraph Eigenvector Centralities}}.
\newblock {\em SIAM Journal on Mathematics of Data Science}, 1(2):293--312,
  2019.

\bibitem{benson2018simplicial}
A.~R. Benson, R.~Abebe, M.~T. Schaub, A.~Jadbabaie, and J.~Kleinberg.
\newblock Simplicial closure and higher-order link prediction.
\newblock {\em Proceedings of the National Academy of Sciences},
  115(48):E11221--E11230, Nov. 2018.

\bibitem{bermond1986strategies}
J.-C. Bermond, C.~Delorme, and J.-J. Quisquater.
\newblock Strategies for interconnection networks: {{Some}} methods from graph
  theory.
\newblock {\em Journal of Parallel and Distributed Computing}, 3(4):433--449,
  1986.

\bibitem{bhuiyan2012guise}
M.~A. Bhuiyan, M.~Rahman, M.~Rahman, and M.~Al~Hasan.
\newblock {{GUISE}}: {{Uniform Sampling}} of {{Graphlets}} for {{Large Graph
  Analysis}}.
\newblock In {\em Proceedings of the 12th {{IEEE International Conference}} on
  {{Data Mining}}}, ICDM'12, pages 91--100, 2012.

\bibitem{casiraghi2017relational}
G.~Casiraghi, V.~Nanumyan, I.~Scholtes, and F.~Schweitzer.
\newblock From {{Relational Data}} to {{Graphs}}: {{Inferring Significant Links
  Using Generalized Hypergeometric Ensembles}}.
\newblock In {\em Social {{Informatics}}}, volume 10540, pages 111--120, 2017.

\bibitem{chee2020constrained}
Y.~M. Chee, T.~Etzion, H.~M. Kiah, A.~Vardy, V.~K. Vu, and E.~Yaakobi.
\newblock Constrained de bruijn codes: {{Properties}}, enumeration,
  constructions, and applications.
\newblock {\em arXiv:2005.03102}, 2020.

\bibitem{chiericetti2016sampling}
F.~Chiericetti, A.~Dasgupta, R.~Kumar, S.~Lattanzi, and T.~Sarl{\'o}s.
\newblock On sampling nodes in a network.
\newblock In {\em Proceedings of the 25th International Conference on World
  Wide Web}, {{WWW}}'16, pages 471--481, 2016.

\bibitem{chikhi2015representation}
R.~Chikhi, A.~Limasset, S.~Jackman, J.~T. Simpson, and P.~Medvedev.
\newblock On the {{Representation}} of {{De Bruijn Graphs}}.
\newblock {\em Journal of Computational Biology}, 22(5):336--352, May 2015.

\bibitem{chodrow2020configuration}
P.~S. Chodrow.
\newblock Configuration models of random hypergraphs.
\newblock {\em Journal of Complex Networks}, 8(3), 2020.

\bibitem{cooper2014estimating}
C.~Cooper, T.~Radzik, and Y.~Siantos.
\newblock Estimating network parameters using random walks.
\newblock {\em Social Network Analysis and Mining}, 4(1):168, Dec. 2014.

\bibitem{cooper2016fast}
C.~Cooper, T.~Radzik, and Y.~Siantos.
\newblock Fast low-cost estimation of network properties using random walks.
\newblock {\em Internet Mathematics}, 12(4):221--238, 2016.

\bibitem{costa2007exploring}
L.~d.~F. Costa and G.~Travieso.
\newblock Exploring complex networks through random walks.
\newblock {\em Physical Review E: Statistical Physics, Plasmas, Fluids, and
  Related Interdisciplinary Topics}, 75(1):016102, Jan. 2007.

\bibitem{dean2008mapreduce}
J.~Dean and S.~Ghemawat.
\newblock {{MapReduce}}: {{Simplified}} data processing on large clusters.
\newblock {\em Communications of the ACM}, 51(1):107--113, Jan. 2008.

\bibitem{erdos1960evolution}
P.~Erd\H{o}s and A.~R{\'e}nyi.
\newblock On the evolution of random graphs.
\newblock {\em Publications of the Mathematical Institute of the Hungarian
  Academy of Sciences}, 5(1):17--60, 1960.

\bibitem{evans2009line}
T.~S. Evans and R.~Lambiotte.
\newblock Line graphs, link partitions, and overlapping communities.
\newblock {\em Physical Review E}, 80(1):016105, July 2009.

\bibitem{faizian2018random}
P.~Faizian, M.~A. Mollah, X.~Yuan, Z.~Alzaid, S.~Pakin, and M.~Lang.
\newblock Random {{Regular Graph}} and {{Generalized De Bruijn Graph}} with
  \$k\$ -{{Shortest Path Routing}}.
\newblock {\em IEEE Transactions on Parallel and Distributed Systems},
  29(1):144--155, Jan. 2018.

\bibitem{garimella2020detection}
K.~V. Garimella, Z.~Iqbal, M.~A. Krause, S.~Campino, M.~Kekre, E.~Drury,
  D.~Kwiatkowski, J.~M. S{\'a}, T.~E. Wellems, and G.~McVean.
\newblock Detection of simple and complex de novo mutations with multiple
  reference sequences.
\newblock {\em Genome Research}, page genome;gr.255505.119v1, Aug. 2020.

\bibitem{gilbert1959random}
E.~N. Gilbert.
\newblock Random graphs.
\newblock {\em The Annals of Mathematical Statistics}, 30(4):1141--1144, 1959.

\bibitem{gkantsidis2006random}
C.~Gkantsidis, M.~Mihail, and A.~Saberi.
\newblock Random walks in peer-to-peer networks: {{Algorithms}} and evaluation.
\newblock {\em Performance Evaluation}, 63(3):241--263, 2006.

\bibitem{gote2020predicting}
C.~Gote, G.~Casiraghi, F.~Schweitzer, and I.~Scholtes.
\newblock Predicting {{Sequences}} of {{Traversed Nodes}} in {{Graphs}} using
  {{Network Models}} with {{Multiple Higher Orders}}.
\newblock {\em arxiv:2007.06662}, July 2020.

\bibitem{hartle2020network}
H.~Hartle, B.~Klein, S.~McCabe, A.~Daniels, G.~{St-Onge}, C.~Murphy, and
  L.~{H{\'e}bert-Dufresne}.
\newblock Network comparison and the within-ensemble graph distance.
\newblock {\em Proceedings of the Royal Society A: Mathematical, Physical and
  Engineering Sciences}, 476(2243):20190744, Nov. 2020.

\bibitem{iacopini2019simplicial}
I.~Iacopini, G.~Petri, A.~Barrat, and V.~Latora.
\newblock Simplicial models of social contagion.
\newblock {\em Nature Communications}, 10(1):2485, Dec. 2019.

\bibitem{iqbal2012novo}
Z.~Iqbal, M.~Caccamo, I.~Turner, P.~Flicek, and G.~McVean.
\newblock De novo assembly and genotyping of variants using colored de
  {{Bruijn}} graphs.
\newblock {\em Nature Genetics}, 44(2):226--232, 2012.

\bibitem{jazayeri2020motif}
A.~Jazayeri and C.~C. Yang.
\newblock Motif discovery algorithms in static and temporal networks: {{A}}
  survey.
\newblock {\em Journal of Complex Networks}, 8(4), Dec. 2020.

\bibitem{jurgens2012temporal}
D.~Jurgens and T.-C. Lu.
\newblock Temporal {{Motifs Reveal}} the {{Dynamics}} of {{Editor
  Interactions}} in {{Wikipedia}}.
\newblock In {\em Proceedings of 2012 the International {{AAAI}} Conference on
  Web and Social Media}, ICWSM'12, 2012.

\bibitem{kovanen2011temporal}
L.~Kovanen, M.~Karsai, K.~Kaski, J.~Kert{\'e}sz, and J.~Saram{\"a}ki.
\newblock Temporal motifs in time-dependent networks.
\newblock {\em Journal of Statistical Mechanics: Theory and Experiment},
  2011(11):P11005, Nov. 2011.

\bibitem{kovanen2013temporal}
L.~Kovanen, K.~Kaski, J.~Kert{\'e}sz, and J.~Saram{\"a}ki.
\newblock Temporal motifs reveal homophily, gender-specific patterns, and group
  talk in call sequences.
\newblock {\em Proceedings of the National Academy of Sciences},
  110(45):18070--18075, 2013.

\bibitem{lambiotte2019networks}
R.~Lambiotte, M.~Rosvall, and I.~Scholtes.
\newblock From networks to optimal higher-order models of complex systems.
\newblock {\em Nature Physics}, 15(4):313--320, 2019.

\bibitem{larock2021debruijn}
T.~LaRock.
\newblock {DeBruijnNets.jl} software package.
\newblock \url{https://www.github.com/tlarock/DeBruijnNets.jl}, 2021.

\bibitem{larock2020hypa}
T.~LaRock, V.~Nanumyan, I.~Scholtes, G.~Casiraghi, T.~{Eliassi-Rad}, and
  F.~Schweitzer.
\newblock {{HYPA}}: {{Efficient Detection}} of {{Path Anomalies}} in {{Time
  Series Data}} on {{Networks}}.
\newblock In {\em Proceedings of the 2020 {{SIAM International Conference}} on
  {{Data Mining}}}, {{SDM}}'20, pages 460--468, 2020.

\bibitem{lee2020hypergraph}
G.~Lee, J.~Ko, and K.~Shin.
\newblock Hypergraph {{Motifs}}: {{Concepts}}, {{Algorithms}}, and
  {{Discoveries}}.
\newblock {\em Proceedings of the VLDB Endowment}, 13(12):2256--2269, Aug.
  2020.

\bibitem{lempel1970homomorphism}
A.~Lempel.
\newblock On a {{Homomorphism}} of the de {{Bruijn Graph}} and its
  {{Applications}} to the {{Design}} of {{Feedback Shift Registers}}.
\newblock {\em IEEE Transactions on Computers}, C-19(12):1204--1209, 1970.

\bibitem{lempel1971extremal}
A.~Lempel.
\newblock On extremal factors of the de {{Bruijn}} graph.
\newblock {\em Journal of Combinatorial Theory, Series B}, 11(1):17--27, Aug.
  1971.

\bibitem{liu2020temporal}
P.~Liu, V.~Guarrasi, and A.~E. Sar{\i}y{\"u}ce.
\newblock Temporal {{Network Motifs}}: {{Models}}, {{Limitations}},
  {{Evaluation}}.
\newblock {\em arxiv:2005.11817}, 2020.

\bibitem{loguinov2005graphtheoretic}
D.~Loguinov, J.~Casas, and X.~Wang.
\newblock Graph-theoretic {{Analysis}} of {{Structured Peer-to-Peer Systems}}:
  {{Routing Distances}} and {{Fault Resilience}}.
\newblock {\em IEEE/ACM Transactions on Networking}, 13(5):1107--1120, 2005.

\bibitem{lu2012sampling}
X.~Lu and S.~Bressan.
\newblock Sampling connected induced subgraphs uniformly at random.
\newblock In {\em Proceeding of the 24th {{International Conference}} on
  {{Scientific}} and {{Statistical Database Management}}}, {{SSDBM}}'12, pages
  195--212, 2012.

\bibitem{milo2002network}
R.~Milo, S.~{Shen-Orr}, S.~Itzkovitz, N.~Kashtan, D.~Chklovskii, and U.~Alon.
\newblock Network {{Motifs}}: {{Simple Building Blocks}} of {{Complex
  Networks}}.
\newblock {\em Science}, 298(5594):824--827, 2002.

\bibitem{mykkeltveit1972proof}
J.~Mykkeltveit.
\newblock A proof of {{Golomb}}'s conjecture for the de {{Bruijn}} graph.
\newblock {\em Journal of Combinatorial Theory, Series B}, 13(1):40--45, Aug.
  1972.

\bibitem{paranjape2017motifs}
A.~Paranjape, A.~R. Benson, and J.~Leskovec.
\newblock Motifs in {{Temporal Networks}}.
\newblock In {\em Proceedings of the {{10th ACM International Conference}} on
  {{Web Search}} and {{Data Mining}}}, {{WSDM}}'17, pages 601--610, 2017.

\bibitem{patra2020review}
S.~Patra and A.~Mohapatra.
\newblock Review of tools and algorithms for network motif discovery in
  biological networks.
\newblock {\em IET Systems Biology}, 14(4):171--189, Aug. 2020.

\bibitem{petri2018simplicial}
G.~Petri and A.~Barrat.
\newblock Simplicial {{Activity Driven Model}}.
\newblock {\em Physical Review Letters}, 121(22):228301, Nov. 2018.

\bibitem{petrovic2019counting}
L.~V. Petrovic and I.~Scholtes.
\newblock Counting {{Causal Paths}} in {{Big Times Series Data}} on
  {{Networks}}.
\newblock {\em arxiv:905.11287}, 2019.

\bibitem{pevzner2001eulerian}
P.~A. Pevzner, H.~Tang, and M.~S. Waterman.
\newblock An {{Eulerian}} path approach to {{DNA}} fragment assembly.
\newblock {\em Proceedings of the National Academy of Sciences},
  98(17):9748--9753, 2001.

\bibitem{pibiri2019handling}
G.~E. Pibiri and R.~Venturini.
\newblock Handling {{Massive N-Gram Datasets Efficiently}}.
\newblock {\em ACM Transactions on Information Systems}, 37(2):1--41, 2019.

\bibitem{ribeiro2010estimating}
B.~Ribeiro and D.~Towsley.
\newblock Estimating and sampling graphs with multidimensional random walks.
\newblock In {\em Proceedings of the 10th {{ACM SIGCOMM}} Conference on
  Internet Measurement}, {{IMC}}'10, pages 390--403, 2010.

\bibitem{ribeiro2012sampling}
B.~Ribeiro, P.~Wang, F.~Murai, and D.~Towsley.
\newblock Sampling directed graphs with random walks.
\newblock In {\em Proceedings of the 2012 {{IEEE INFOCOM}}}, pages 1692--1700,
  2012.

\bibitem{ribeiro2019survey}
P.~Ribeiro, P.~Paredes, M.~E.~P. Silva, D.~Aparicio, and F.~Silva.
\newblock A {{Survey}} on {{Subgraph Counting}}: {{Concepts}}, {{Algorithms}}
  and {{Applications}} to {{Network Motifs}} and {{Graphlets}}.
\newblock {\em arXiv:1910.13011}, Oct. 2019.

\bibitem{ribeiro2014gtries}
P.~Ribeiro and F.~Silva.
\newblock G-{{Tries}}: {{A}} data structure for storing and finding subgraphs.
\newblock {\em Data Mining and Knowledge Discovery}, 28(2):337--377, Mar. 2014.

\bibitem{rossi2019heterogeneous}
R.~A. Rossi, N.~K. Ahmed, A.~Carranza, D.~Arbour, A.~Rao, S.~Kim, and E.~Koh.
\newblock Heterogeneous {{Network Motifs}}.
\newblock {\em arxiv:1901.10026}, Jan. 2019.

\bibitem{saramaki2005characterizing}
J.~Saram{\"a}ki.
\newblock Characterizing {{Motifs}} in {{Weighted Complex Networks}}.
\newblock In {\em {{AIP Conference Proceedings}}}, volume 776, pages 108--117.
  {AIP}, 2005.

\bibitem{sarkar2019using}
S.~Sarkar, R.~Guo, and P.~Shakarian.
\newblock Using network motifs to characterize temporal network evolution
  leading to diffusion inhibition.
\newblock {\em Social Network Analysis and Mining}, 9(1):14:1--14:24, Dec.
  2019.

\bibitem{scholtes2017when}
I.~Scholtes.
\newblock When is a network a network?: {{Multi-order}} graphical model
  selection in pathways and temporal networks.
\newblock In {\em Proceedings of the 23rd {{ACM SIGKDD}} International
  Conference on Knowledge Discovery and Data Mining}, {{KDD}}'17, pages
  1037--1046, 2017.

\bibitem{scholtes2016higherorder}
I.~Scholtes, N.~Wider, and A.~Garas.
\newblock Higher-order aggregate networks in the analysis of temporal networks:
  Path structures and centralities.
\newblock {\em The European Physical Journal B}, 89(3), Mar. 2016.

\bibitem{scholtes2014causalitydriven}
I.~Scholtes, N.~Wider, R.~Pfitzner, A.~Garas, C.~J. Tessone, and F.~Schweitzer.
\newblock Causality-driven slow-down and speed-up of diffusion in
  non-{{Markovian}} temporal networks.
\newblock {\em Nature Communications}, 5(1), Dec. 2014.

\bibitem{schwarze2020motifs}
A.~C. Schwarze and M.~A. Porter.
\newblock Motifs for processes on networks.
\newblock {\em arxiv:2007.07447}, 2020.

\bibitem{sekara2016fundamental}
V.~Sekara, A.~Stopczynski, and S.~Lehmann.
\newblock Fundamental structures of dynamic social networks.
\newblock {\em Proceedings of the National Academy of Sciences},
  113(36):9977--9982, 2016.

\bibitem{shen-orr2002network}
S.~S. {Shen-Orr}, R.~Milo, S.~Mangan, and U.~Alon.
\newblock Network motifs in the {{Transcriptional}} regulation network of
  {{Escherichia}} coli.
\newblock {\em Nature genetics}, 31(1):64, 2002.

\bibitem{sinatra2010networks}
R.~Sinatra, D.~Condorelli, and V.~Latora.
\newblock Networks of {{Motifs}} from {{Sequences}} of {{Symbols}}.
\newblock {\em Physical Review Letters}, 105(17):178702, Oct. 2010.

\bibitem{soundarajan2016generating}
S.~Soundarajan, A.~Tamersoy, E.~B. Khalil, T.~{Eliassi-Rad}, D.~H. Chau,
  B.~Gallagher, and K.~Roundy.
\newblock Generating {{Graph Snapshots}} from {{Streaming Edge Data}}.
\newblock In {\em Proceedings of the 25th {{International World Wide Web
  Conference}}}, {{WWW}}'16, pages 109--110, 2016.

\bibitem{stone2019network}
L.~Stone, D.~Simberloff, and Y.~{Artzy-Randrup}.
\newblock Network motifs and their origins.
\newblock {\em PLOS Computational Biology}, 15(4):e1006749, Apr. 2019.

\bibitem{torres2021why}
L.~Torres, A.~S. Blevins, D.~Bassett, and T.~{Eliassi-Rad}.
\newblock The {{Why}}, {{How}}, and {{When}} of {{Representations}} for
  {{Complex Systems}}.
\newblock {\em SIAM Review}, 63(3):435--485, 2021.

\bibitem{transstat2018origin}
R.~TransStat.
\newblock Origin and destination survey database, 2018.

\bibitem{tu2018network}
K.~Tu, J.~Li, D.~Towsley, D.~Braines, and L.~D. Turner.
\newblock Network {{Classification}} in {{Temporal Networks Using Motifs}}.
\newblock {\em arXiv:1807.03733}, Aug. 2018.

\bibitem{ugander2013subgraph}
J.~Ugander, L.~Backstrom, and J.~Kleinberg.
\newblock Subgraph {{Frequencies}}: {{Mapping}} the {{Empirical}} and
  {{Extremal Geography}} of {{Large Graph Collections}}.
\newblock In {\em Proceedings of the 22nd {{International Conference}} on
  {{World Wide Web}}}, {{WWW}}'13, pages 1307--1318, 2013.

\bibitem{underwood2020motifbased}
W.~G. Underwood, A.~Elliott, and M.~Cucuringu.
\newblock Motif-based spectral clustering of weighted directed networks.
\newblock {\em Applied Network Science}, 5(1):62, Dec. 2020.

\bibitem{wasserman1994social}
S.~Wasserman and K.~Faust.
\newblock {\em Social Network Analysis: {{Methods}} and Applications}.
\newblock {Cambridge University Press}, 1994.

\bibitem{west2012human}
R.~West and J.~Leskovec.
\newblock Human wayfinding in information networks.
\newblock In {\em Proceedings of the 21st International Conference on {{World
  Wide Web}}}, WWW'12, pages 619--628, 2012.

\bibitem{xu2016representing}
J.~Xu, T.~L. Wickramarathne, and N.~V. Chawla.
\newblock Representing higher-order dependencies in networks.
\newblock {\em Science Advances}, 2(5):e1600028--e1600028, 2016.

\bibitem{xuan2015temporal}
Q.~Xuan, H.~Fang, C.~Fu, and V.~Filkov.
\newblock Temporal motifs reveal collaboration patterns in online task-oriented
  networks.
\newblock {\em Physical Review E: Statistical Physics, Plasmas, Fluids, and
  Related Interdisciplinary Topics}, 91(5):052813, May 2015.

\bibitem{yan2003closegraph}
X.~Yan and J.~Han.
\newblock {{CloseGraph}}: {{Mining Closed Frequent Graph Patterns}}.
\newblock In {\em Proceedings of the {{9th ACM SIGKDD International
  Conference}} on {{Knowledge Discovery}} and {{Data Mining}}}, {{KDD}}'03,
  pages 286--295, 2003.

\bibitem{yan2005mining}
X.~Yan, X.~J. Zhou, and J.~Han.
\newblock Mining closed relational graphs with connectivity constraints.
\newblock In {\em Proceeding of the 11th {{ACM SIGKDD}} International
  Conference on {{Knowledge}} Discovery in Data Mining}, KDD'05, pages
  324--333, 2005.

\bibitem{yoon2007statistical}
S.~Yoon, S.~Lee, S.-H. Yook, and Y.~Kim.
\newblock Statistical properties of sampled networks by random walks.
\newblock {\em Physical Review E: Statistical Physics, Plasmas, Fluids, and
  Related Interdisciplinary Topics}, 75(4):046114, Apr. 2007.

\bibitem{young2017construction}
J.-G. Young, G.~Petri, F.~Vaccarino, and A.~Patania.
\newblock Construction of and efficient sampling from the simplicial
  configuration model.
\newblock {\em Physical Review E}, 96(3):032312, 2017.

\bibitem{zerbino2008velvet}
D.~R. Zerbino and E.~Birney.
\newblock Velvet: {{Algorithms}} for de novo short read assembly using de
  {{Bruijn}} graphs.
\newblock {\em Genome Research}, 18(5):821--829, Feb. 2008.

\end{thebibliography}
